\pdfoutput=1
\documentclass[11pt,onecolumn,draftclsnofoot]{IEEEtran}
\usepackage{verbatim}
\usepackage{amsfonts,amsmath,mathrsfs}
\usepackage[final]{graphicx}
\usepackage{times}
\usepackage{psfrag}
\usepackage{amssymb,amsbsy}
\begin{document}
\newtheorem{theorem}{Theorem}
\newtheorem{corrly}{Corollary}
\newtheorem{prop}{Proposition}
\newtheorem{lemma}{Lemma}
\newtheorem{definition}{Definition}
\newtheorem{remark}{Remark}
\newcommand{\entropy}{\mathsf{h}}
\newcommand{\logfn}[1]{\frac{1}{2}\log\left( #1 \right)}
\newcommand{\twobyn}{\frac{2}{n}}
\newcommand{\nbytwo}{\frac{n}{2}}

\newcommand{\prob}{\textsf{P}}
\newcommand{\re}{\mbox{$I\!\!R$}}
\newcommand{\cov}[1]{\text{Cov}\left(#1\right)}
\newcommand{\expect}[1]{\textsf{E}\left[#1\right]}
\newcommand{\Csum}{\mathcal{C}_{\text{sum}}}
\newcommand{\snr}{P}
\newcommand{\BCbound}{\text{broadcast channel outer bound}}
\newcommand{\ZCbound}{\text{Z-channel sum rate outer bound}}
\newcommand{\muser}{$M$--user }
\newcommand{\gic}{Gaussian interference channel }
\newcommand{\onemany}{one-to-many }
\newcommand{\manyone}{many-to-one }
\newcommand{\uy}{\underline{Y}}
\newcommand{\uyt}{\underline{\tilde{Y}}}
\newcommand{\ux}{\underline{X}}
\newcommand{\uxt}{\underline{\tilde{X}}}
\newcommand{\uxh}{\underline{\hat{X}}}
\newcommand{\uz}{\underline{Z}}
\newcommand{\uzt}{\underline{\tilde{Z}}}
\newcommand{\uzh}{\underline{\hat{Z}}}
\newcommand{\us}{\underline{S}}
\newcommand{\hs}{\hat{S}}
\newcommand{\ush}{\underline{\hat{X}}}
\newcommand{\ust}{\underline{\tilde{S}}}
\newcommand{\uh}{\underline{h}}
\newcommand{\uv}{\underline{V}}
\newcommand{\uw}{\underline{W}}
\newcommand{\uzero}{\underline{0}}
\newcommand{\bbH}{\mathbb{H}}
\newcommand{\bbA}{\mathbb{A}}
\newcommand{\bbC}{\mathbb{B}}
\newcommand{\chanmatrix}{\bbH}


\title{{\small Submitted to IEEE Transactions on Information Theory, February 2008. Revised Nov 2008.}\\Gaussian Interference Networks: Sum Capacity in the Low Interference Regime and New Outer Bounds on the Capacity Region}
\author{
{\large{V. Sreekanth Annapureddy and Venugopal V. Veeravalli$^{*}$} \\ {\texttt{\{vannapu2,vvv\}@uiuc.edu.}}}
\thanks{$^*$ The authors are with the Coordinated Science Laboratory and the
Department of Electrical and Computer Engineering,
University of Illinois at Urbana-Champaign, Urbana, IL 61801 USA.}
\thanks{This research was supported in part by the NSF award CCF 0431088, through the University of Illinois, by a Vodafone Foundation Graduate Fellowship, and a grant from Texas Instruments.}
\thanks{This paper was presented in part at the Information Theory and Applications (ITA) workshop, UCSD, San Diego CA, January 2008 \cite{ITA2008} and at the International Symposium on Information Theory (ISIT), Toronto, Canada, July 2008 \cite{ISIT2008}.}
}
\maketitle
\begin{abstract}
Establishing the capacity region of a Gaussian interference network is an open problem in information theory. Recent progress on this problem has led to the characterization of the capacity region of a general two-user Gaussian interference channel within one bit. In this paper, we develop new, improved  outer bounds on the capacity region. Using these  bounds, we show that {\em treating interference as noise} achieves the {\em sum capacity} of the two-user Gaussian interference channel in a {\em low interference regime}, where the interference parameters are below certain thresholds. We then generalize our techniques and results to Gaussian interference networks with more than two users. In particular, we demonstrate that the total interference threshold, below which treating interference as noise achieves the sum capacity, increases with the number of users.
\end{abstract}

\section{Introduction}
In his celebrated paper \cite{Shannon1948}, Shannon established the capacity of the additive white Gaussian noise (AWGN) channel, where the performance is limited by thermal noise. In multiuser wireless networks, the performance is also limited by the interference from other users sharing the same spectrum. Unlike thermal noise, interference has a definite structure since it is generated by other users. Can this structure be exploited to decrease the uncertainty and thus improve the performance of the communication network? If so, what are the optimal signaling strategies? In this paper, we establish the somewhat counter-intuitive result that exploiting the structure of the interference in Gaussian interference channels does not improve the overall system throughput in a {\em low interference} regime. In other words, it is possible to {\em treat interference as noise} and still achieve the maximum possible throughput, if the interference levels are below certain thresholds.
\begin{figure}[htb!]
\centering
\includegraphics[width=.5\textwidth]{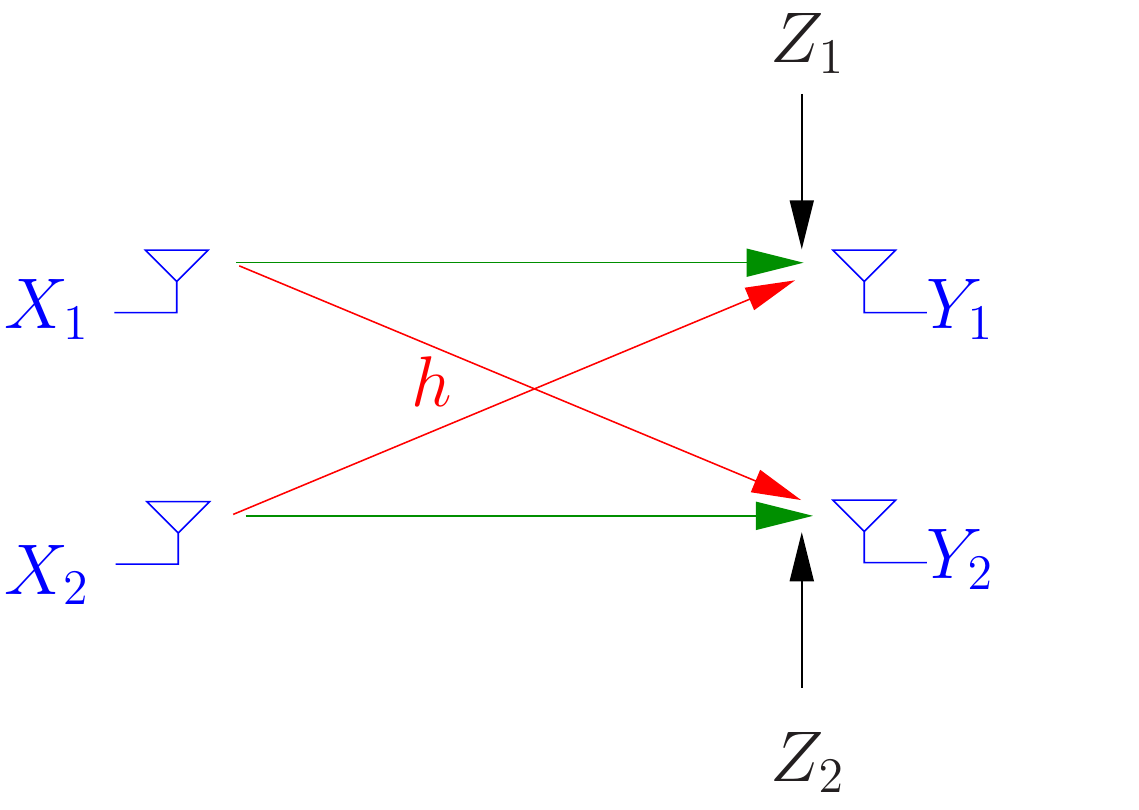}
\caption{Two-user symmetric Gaussian interference channel}
\label{fig:twouser-gic}
\end{figure}

%
Interference management is of vital importance in wireless communication systems, with several users contending for the same limited spectrum.
%
%
As a first step towards an information-theoretic study of interference management, consider two users sharing a wireless channel as shown in Figure \ref{fig:twouser-gic}, where each user's receiver is interested in only the information transmitted by the corresponding transmitter. Each user's rate of communication is limited by the Gaussian noise at the receiver and the interference caused by the other user. Carleial \cite{Carleial1975} showed that interference does not reduce the capacity of such a {\em two-user Gaussian interference channel} in the {\em very strong interference} setting, where each receiver can completely cancel the interference by exploiting its structure. Subsequently, the capacity region was determined in the {\em strong interference} setting \cite{HK1981, Sato1981}, where it was shown that each user can decode the information transmitted to the other user. Establishing the capacity region in the other regimes remains an open problem. The best known achievable region for the two-user Gaussian interference channel is based on the Han-Kobayashi (HK) scheme \cite{HK1981,CMG2007}. Here the users split message into private and common messages, and each user jointly decodes its own messages and the common message of the interfering user. This is in general a sophisticated scheme, requiring  multi-user encoders and decoders and coordination between the users. What we establish in this paper is that if the interference levels are low enough, then the receivers can treat interference as noise, and single users encoders and decoders can be employed without any loss in sum capacity.

In order to establish the sum capacity in the {\em low interference} regime, we need to prove a converse, i.e., derive an outer bound on the sum capacity that matches with the sum rate achieved by treating interference as noise. The concept of a {\em genie} giving side information to the receivers was used in \cite{Kramer2004,OneBit2007} to derive outer bounds on the capacity region. Since the receivers can choose not to use the side information, the capacity region of the genie-aided channel is an obvious outer bound to the capacity region of the interference channel. In \cite{OneBit2007}, a specific set of genie-aided outer bounds are shown to be  {\em within one bit} of the capacity region. We show that the bounding technique developed in \cite{OneBit2007} is applicable to a wider class of genie signals. We further show that if the channel parameters satisfy a condition for low interference, the genie can be selected in a clever way so that the resulting genie-aided outer bound matches the sum rate achievable by treating interference as noise. With this wider class of genie signals and using the entropy power inequality \cite{cover-thomas}, we also derive outer bounds on the entire capacity region that are tighter than existing outer bounds. Similar results have also been established independently by Shang et. al in \cite{Shang-Kramer-Chen-2007} and Motahari et. al. in \cite{Motahari-Khandani-2007}.

We then generalize the results to Gaussian interference networks with more than two users. Using a genie similar to that used for the two-user channel, we derive low interference regime conditions for the many-to-one interference channel, where the interference is experienced by only one user and one-to-many interference channel, where the interference is generated by only one user. We also propose a new genie construction, where each receiver is provided with multiple genie signals, for any arbitrary Gaussian interference network. This genie is a generalization of the genie used in \cite{OneBit2007} and the purpose of this generalization is to develop results analogous to \cite{OneBit2007} for arbitrary Gaussian interference networks. We show that treating interference as noise with Gaussian inputs achieves the sum capacity of the vector genie-aided channel. As done for the two-user channel, this outer bound can be tightened to establish the sum capacity in a low interference regime. We tighten the bound for a three-user symmetric Gaussian interference channel, and demonstrate the existence of channels for which treating interference as noise is optimal, but the total interference to noise ratio (INR) is greater than the INR threshold of the two-user interference channel.



\subsection{Notation and Organization}
We use the following notation. For deterministic objects, we use lowercase letters for scalars and uppercase letters in blackboard bold font for matrices. For example, we use $h$ to denote a deterministic scalar and $\chanmatrix{}$ to denote a deterministic matrix. For random objects, we use uppercase letters for scalars, and underlined uppercase letters for vectors. Random objects with superscripts denote sequences of the random objects in time. For example, we use $X$ to denote a random scalar, $\ux$ to denote a random vector, and $X^n$ and $\ux^n$ to denote the sequences of length $n$ of the random scalars and vectors, respectively. We use $\cov{X}$ to denote the variance of a random variable $X$, and $\cov{X|Y}$ denote the minimum mean square error in estimating the random variable $X$ from the random variable $Y$, with similar notation for random vectors. We use $\mathcal {N} (\mu,\sigma^2)$ to denote the Gaussian distribution with mean $\mu$ and variance $\sigma^2$, and $\mathcal {N} (\underline{\mu},\Sigma)$ to denote the Gaussian vector distribution with mean $\underline{\mu}$ and covariance matrix $\Sigma$. We use $\mathsf{h}(.)$ to denote the differential entropy of a continuous random variable or vector and $I(.;.)$ to denote the mutual information.

The rest of the paper is organized as follows. In section~\ref{Sec:channel-model}, we introduce the model for the Gaussian interference network that we study. In Section~\ref{Sec:mathematical-prelims}, we summarize mathematical results such as the entropy power inequality and prove some new results that are required in establishing our new outer bounds. In Section~\ref{Sec:twouser-previouswork}, we review the existing bounds on the capacity region of the two-user Gaussian interference channel. In Section~\ref{Sec:twouser-sumcapacity}, we establish the sum capacity of two-user Gaussian interference channel in a low interference regime. In Section~\ref{Sec:twouser-outerbounds-capacityregion}, we present new outer bounds on the capacity region of the two-user channel. In Section~\ref{Sec:network-sumcapacity}, we present extensions of our results on the sum capacity  in the low interference regime to Gaussian interference networks with more than two users. In Section~\ref{sec:concl}, we provide some concluding remarks.
\section{Interference Network Model}
\label{Sec:channel-model}
Consider a Gaussian interference network with $M$ users, i.e., $M$ pairs of transmitters and receivers, where no user is interested in the information transmitted to the other users. Over one symbol period, the channel is described by
 \begin{equation}
Y_r = \sum_{t = 1}^{M} h_{rt}X_t + Z_r, 1 \leq r \leq M
\label{channel_output_rth_rx}
\end{equation}
where $X_t$ is the signal transmitted by transmitter $t$, $h_{rt}$ is the fixed channel gain from transmitter $t$ to receiver $r$, and the receiver noise terms $\{Z_r\}_{r=1}^{M}$ are assumed to be zero mean, unit variance, independent Gaussian random variables. Furthermore, the noise is assumed to be independent and identically distributed (i.i.d.) in time. Transmitter $t$ has an average power constraint $\snr_t$. In vector notation, \eqref{channel_output_rth_rx} is equivalent to
 \begin{equation}
\uy = \chanmatrix{}\ux + \uz
\label{interference_network}
\end{equation}
where $\chanmatrix{}$ is a deterministic $M\times M$-matrix with elements $\{h_{r,t}\}$.

The interference network is said to be in standard form  \cite{Carleial1978}, if
\[
h_{rt} = 1, \forall r = t.
\]
Any interference network \eqref{interference_network} can be expressed in an equivalent standard form for the purposes of an information-theoretic analysis. For each user $i$, let the message index $m_i$ be uniformly distributed over $\{1,2,\ldots,2^{nR_i}\}$, and let $\mathcal{C}_i(n)$ be a code consisting of an encoding function $X_i^n: \{1,2,\ldots,2^{nR_i}\} \rightarrow \re^n$ satisfying the power constraint
\[
||X_i^n(m_i)||^2 \leq nP_i, \forall m_i \in \{1,2,\ldots,2^{nR_i}\}
\]
and a decoding function $g_i: \re^n \rightarrow \{1,2,\ldots,2^{nR_i}\}$. The corresponding probability of decoding error $\lambda_i(n)$ is defined as $\prob\{ m_i \neq g_i(Y_i^n)\}$. A rate tuple $(R_1,R_2, \ldots , R_M)$ is said to be achievable if there exists a sequence of codes $\{\mathcal{C}_1(n),\mathcal{C}_2(n), \ldots, \mathcal{C}_M(n)\}_{n=1}^{\infty}$ such that the error probabilities $\lambda_1(n), \lambda_2(n), \ldots,$ and $\lambda_M(n)$ all go to zero as $n$ goes to infinity. Capacity region is the closure of all the achievable rate tuples.
\section{Mathematical Preliminaries}
\label{Sec:mathematical-prelims}
In this section, we review the information inequalities that are useful in establishing our new outer bounds.
%
%
The first result is a generalization of the maximum entropy theorem. Consider a sequence of random variables $\{X_j\}_{j=1}^{n}$ with average power constraint $\sum_{j=1}^{n}\expect{X_j^2} \leq nP$.  It is well known that $\entropy(X^n) \leq \frac{n}{2}\log(2\pi{}eP)$, and equality is achieved if and only if (iff) $\{X_j\}_{j=1}^{n}$ are i.i.d. $\mathcal{N}(0,P)$  \cite[Theorem 8.6.5]{cover-thomas}. The following lemma is a generalization of this result.
\begin{lemma}
Let $\ux$ be a random vector, and let $\uy$ and $\us$ be noisy observations of $\ux$.
\[
\begin{split}
\uy & =  \bbA \; \ux + \uz \\
\us & =  \bbC \; \ux + \uw
\end{split}
\]
where $\uz$ and $\uw$ are correlated, zero-mean, Gaussian random vectors, and $\bbA$ and $\bbC$ are real valued matrices. Consider the random vector sequence $\ux^n = (\ux_1, \ldots, \ux_n)$ with the covariance constraint $\frac{1}{n}\sum_{j=1}^{n} \Sigma_{xj} \preceq \Sigma_x$, where $\Sigma_{xj}$ is the covariance matrix of $\ux_j$. Furthermore, let $\uy^n$ and $\us^n$ be the corresponding observations when the noise vector sequences $\uz^n$ and $\uw^n$ each have components that are i.i.d. in time. Then, we have
\[
\entropy(\uy^n|\us^n) \leq n\entropy(\uy_G|\us_G)
\]
where $\uy_G$ and $\us_G$ are $\uy$ and $\us$ when $\ux = \ux_G \sim \mathcal{N}(\uzero,\Sigma_x)$.
\label{lemma-concavity-conditonal-entropy}
\end{lemma}
\begin{proof}
Let $Q$ be a time sharing random variable taking values from $1$ to $n$ with equal probability. Let $\uxt_G \sim \mathcal{N}(0,\frac{1}{n}\sum_{i=1}^{n}\Sigma_{xi})$, and $\uyt_G$ and $\ust_G$ be the corresponding $\uy$ and $\us$.
\[
\begin{split}
\entropy(\uy^n|\us^n) = & \ \sum_{i = 1}^{n} \entropy(\uy_i|\uy^{i-1},\us^n) \\
\stackrel{(a)} \leq & \ \sum_{i = 1}^{n} \entropy(\uy_i|\us_i) \\
= & \ n\entropy(\uy_Q|\us_Q,Q) \\
\stackrel{(b)} \leq & \ n\entropy(\uy_Q|\us_Q) \\
\stackrel{(c)} \leq & \ n\entropy(\uyt_{G}|\ust_{G}) \\
\end{split}
\]
where the steps (a), (b) follow from the fact that conditioning reduces entropy and step (c) follows because Gaussian distribution maximizes the conditional distribution for a given covariance constraint~\cite[Lemma~1]{Thomas-IT1987}.

Now letting $\uxh_G \sim \mathcal{N}(0,\Sigma_x\; -\; \frac{1}{n}\sum_{i=1}^{n}\Sigma_{xi})$, and further assuming that $\uxh_G$ is independent of $\uxt_G$, $\uz$ and $\uw$, we have
\[
\begin{split}
\entropy(\uyt_{G}|\ust_{G})
= & \ \entropy(\bbA\; \uxt_G + \uz|\bbC \;\uxt_G + \uw) \\
= & \ \entropy(\bbA\; (\uxt_G + \uxh_G ) + \uz|\bbC (\uxt_G + \uxh_G ) + \uw,\uxh_G ) \\
\stackrel{(d)} \leq & \ \entropy(\bbA\; (\uxt_G + \uxh_G ) + \uz|\bbC\;  (\uxt_G + \uxh_G ) + \uw) \\
= & \ \entropy(\uy_G|\us_G)
\end{split}
\]
where the step (d) follow from the fact that conditioning reduces entropy.
\end{proof}

The following is the celebrated entropy power inequality (EPI) \cite[Theorem 17.7.3]{cover-thomas} originally proposed by Shannon.
\begin{lemma}[EPI]
\label{Lemma:EPI}
For any independent random sequences $X^n$ and $Z^n$,
\[
2^{\twobyn{}\entropy(X^n+Z^n)} \geq 2^{\twobyn{}\entropy(X^n)} + 2^{\twobyn{}\entropy(Z^n)}.
\]
\end{lemma}
Often, we are interested in the case where the sequence $Z^n$ is i.i.d. Gaussian, in which case we have the following corollary.
\begin{corrly}
Let $X^n$ be a random sequence and $Z^n$ be an independent random sequence with components that are i.i.d. ${\cal N}(0,\sigma^2)$. Then
\[
\entropy(X^n+Z^n) \geq \frac{n}{2} \log\left(2^{\twobyn{}\entropy(X^n)} + 2\pi{}e\sigma^2\right).
\]
Equivalently
\[
\entropy(X^n) \leq \frac{n}{2} \log\left(2^{\twobyn{}\entropy(X^n+Z^n)} - 2\pi{}e\sigma^2\right).
\]
\label{corollary-EPI}
\end{corrly}
As a corollary of the EPI, we have the worst case noise result that says that if the input distribution is i.i.d. Gaussian, then the noise that minimizes the mutual information under an average power constraint is also i.i.d. Gaussian. (See the mutual information game problem: 9.21 in \cite{cover-thomas}.) With a little abuse of notation, the worst case noise results in the scalar and vector cases are as follows:
\begin{lemma}[Worst Case Noise: Scalar Case]
Let $X^n$ be a random sequence with average power constraint $P$, i.e., $\sum_{j=1}^{n}\expect{X_j^2} \leq nP$, and let $Z^n$ be an independent random sequence with components that are i.i.d. ${\cal N} (0,\sigma^2)$. Then
\[
\entropy(X^n) - \entropy(X^n+Z^n) \leq n\entropy(X_G) - n\entropy(X_G+Z)
\]
where $X_G \sim \mathcal{N}(0,P)$, and equality is achieved if $X^n = X_G^n$, where $X_G^n$ denotes the random sequence with components that are i.i.d. ${\cal N} (0,P)$.
\label{lemma-worstcase-noise-scalar}
\end{lemma}
\begin{proof}
The result follows from the EPI (see proof of Lemma~\ref{new-extremal-inequality} below); a different proof is given in \cite{Diggavi2001}. Interestingly, the result can be established as a direct consequence of the Lemma~\ref{lemma-concavity-conditonal-entropy}, as seen below in  the proof of the Lemma~\ref{lemma-worstcase-noise-vector}.
\end{proof}
\begin{lemma}[Worst Case Noise: Vector Case]
Let $\ux^n$ be a random vector sequence with an average covariance constraint, i.e., $\sum_{j=1}^{n}\Sigma_{xj} \preceq n\Sigma_x$, and let $\uz^n$ be an independent random vector sequence, with components that are i.i.d. ${\cal N} (\uzero, \Sigma_z)$. Then
\[
\entropy(\ux^n) - \entropy(\ux^n+\uz^n) \leq n\entropy(\ux_G) - n\entropy(\ux_G+\uz)
\]
where $\ux_G \sim {\cal N} (\uzero,\Sigma_x)$, and equality is achieved if $\ux^n = \ux_G^n$, where $\ux_G^n$ denotes the random sequence with components that are i.i.d. ${\cal N} (\uzero, \Sigma_x)$.
\label{lemma-worstcase-noise-vector}
\end{lemma}
\begin{proof}
Although the proof follows from  results given in  \cite{Diggavi2001}, we provide a different simple proof based on Lemma~\ref{lemma-concavity-conditonal-entropy}.
\[
\begin{split}
\entropy(\ux^n) - \entropy(\ux^n+\uz^n) = & \ -I(\uz^n;\ux^n+\uz^n) \\
                          = & \ -\entropy(\uz^n) + \entropy(\uz^n|\ux^n+\uz^n) \\
                          = & \ -n\entropy(\uz) + \entropy(\uz^n|\ux^n+\uz^n) \\
        \stackrel{(a)} \leq & \ -n\entropy(\uz) + n\entropy(\uz|\ux_G+\uz) \\
                          = & \ n\entropy(\ux_G) - n\entropy(\ux_G+\uz)
\end{split}
\]
where step (a) follows from Lemma~\ref{lemma-concavity-conditonal-entropy}.
\end{proof}

\begin{remark}
As we have noted in Lemma~\ref{lemma-worstcase-noise-scalar}, the scalar case of the worst case noise result is a corollary of the EPI. However, in the vector case, Lemma~\ref{lemma-worstcase-noise-vector} does not follow from the EPI, unless $\Sigma_x$ is a scaled version of $\Sigma_z$.
\end{remark}

We now provide an extension of the scalar version of the worst case noise result, which is useful in deriving outer bounds on the sum capacity of interference networks with more than two users. This result might also  be useful in other multiuser information theory problems.
\begin{lemma}
For $i = 1,2, \ldots, M$, let $X_i^n$ be a random sequence with average power constraint $P_i$, i.e., $\sum_{j=1}^{n}\expect{X_{ij}^2} \leq nP_i$. Further, let $Z^n$ be a sequence with components that are i.i.d. ${\cal N} (0,\sigma^2)$. Assume that the sequences $X_i^n$ are independent of each other and also independent of $Z^n$,  and let $X_{iG} \sim {\cal N} (0,P_i)$. Then
\begin{equation} \label{eq:newee}
\sum_{i=1}^{M}\lambda_i\entropy(X_i^n) - h\left(\sum_{i=1}^{M}X_i^n+Z^n\right) \leq n\sum_{i=1}^{M}\lambda_i\entropy(X_{iG}) - nh\left(\sum_{i=1}^{M}X_{iG}+Z\right)
\end{equation}
for all
\[
\lambda_i \geq \frac{P_i}{\sum_{i=1}^{M}P_i+\sigma^2}
\]
and equality is achieved in \eqref{eq:newee} if for $i=1,\ldots, M$, $X_i^n = X_{iG}^n$, where $X_{iG}^n$ denotes the random sequence with components that are i.i.d. ${\cal N} (0,P_i)$.
\label{new-extremal-inequality}
\end{lemma}
\begin{proof}
We will prove the lemma for
\[
\lambda_i = \frac{P_i}{\sum_{i=1}^{M}P_i+\sigma^2}
\]
The result with \[\lambda_i > \frac{P_i}{\sum_{i=1}^{M}P_i+\sigma^2}\] follows because the additional positive entropy quantities are easily seen to be maximized by $X_{iG}^n$.

Denote $\frac{\entropy(X_i^n)}{n}$ by $t_i$ and $2\pi{}e\sigma^2$ by $c$. Using the EPI (Lemma~\ref{Lemma:EPI}), we have
\[
\begin{split}
\sum_{i=1}^{M}\lambda_i\entropy(X_i^n) - h\left(\sum_{i=1}^M X_i^n + Z^n\right) \leq & \ n\sum_{i=1}^{M}\lambda_it_i - n\logfn{\sum_{i=1}^M 2^{2t_i} + c} .\\
\end{split}
\]
Let $f(\underline{t}) = \sum_{i=1}^{M}\lambda_it_i - \logfn{\sum_{i=1}^M 2^{2t_i} + c}$. The concavity of $f$ in $\underline{t}$ follows from the convexity of the {\em log-sum-exp} function \cite{boyd-convex-optimization}. Now, using
\[
\frac{\partial f}{\partial t_i} = \lambda_i - \frac{2^{2t_i}}{\sum_{i=1}^M 2^{2t_i} + c}
\]
it can be easily checked that $\{t_j = \logfn{2\pi{}eP_j}\}_{j=1}^M$ satisfy $\frac{\partial f}{\partial t_i} = 0$ for all $i$. Thus, $\{t_j = \logfn{2\pi{}eP_j}\}_{j=1}^M$ maximizes the function $f(\underline{t})$, and hence
\[
\begin{split}
\sum_{i=1}^{M}\lambda_i\entropy(X_i^n) - h\left(\sum_{i=1}^M X_i^n + Z^n\right) \leq & \ nf(\underline{t}) \\
                \leq & \ n\sum_{i=1}^{M}\lambda_i\logfn{2\pi{}eP_i} - n\logfn{2\pi{}e\sum_{i=1}^M P_i + 2\pi{}e\sigma^2} \\
                = & \ n\sum_{i=1}^{M}\lambda_i\entropy(X_{iG}) - nh\left(\sum_{i=1}^{M}X_{iG}+Z\right).
\end{split}
\]
\end{proof}

We now prove the following straightforward lemma, which is nevertheless useful in handling the side information provided by the genie in our genie-aided outer bounds.
\begin{lemma}
Let $\ux^n$ be a random vector sequence,  and let $\uz^n$ and $\uw^n$ be (possibly correlated) zero-mean Gaussian random vector sequences, independent of $\ux^n$ and i.i.d. in time. Then
\[
\entropy(\ux^n + \uz^n|\uw^n) = \entropy(\ux^n + \uv^n)
\]
where $\uv^n$ is i.i.d. $ \mathcal{N}\left(0,\cov{\uz|\uw}\right)$. 
\label{lemma-conditonal-entropy-mmse}
\end{lemma}
\begin{proof}
Let $\hat{\uz}^n$ be the MMSE estimate of $\uz^n$ given $\uw^n$. Then we have
\[
\uz^n = \hat{\uz}^n + \uv^n .
\]
Now
\[
\begin{split}
\entropy(\ux^n + \uz^n|\uw^n) = & \ \entropy(\ux^n + \hat{\uz}^n + \uv^n|\uw^n) \\
      \stackrel{(a)}   = & \ \entropy(\ux^n + \uv^n|\uw^n) \\
      \stackrel{(b)}   = & \ \entropy(\ux^n + \uv^n) \\
\end{split}
\]
where the step (a) follows because the MMSE estimate $\hat{\uz}^n$ is a function of $\uw^n$, and the step (b) follows because the (observation) $\uw^n$ is independent of the MMSE error $\uv^n$ and $\ux^n$.
\end{proof}

\begin{lemma}
For any random vectors $\ux$, $\uy$ and $\us$, 
\begin{enumerate}
\item $I(\ux;\us|\uy) = 0$ iff $\ux - \uy - \us$ form a Markov chain.
\item $\ux - \uy - \us$ form a Markov chain iff $\hat{\us}(\ux,\uy)$, the MMSE estimate of $\us$ given $(\ux,\uy)$, is equal to $\hat{\us}(\uy)$, the MMSE estimate of $\us$ given $\uy$.
\item Furthermore if $X$, $Y$ and $S$ are Gaussian random variables such that 
\[
\begin{split}
Y = & \ X + Z \\
S = & \ X + N
\end{split}
\]
where the zero mean Gaussian random variables $Z$ and $N$ are independent of $X$, then $X - Y -S$ form a Markov chain iff $\expect{NZ}~=~\expect{Z^2}$.
\end{enumerate}
\label{lemma-markov-chain1}
\end{lemma}
\begin{proof}
\begin{enumerate}
\item Claim 1 follows from Theorem~2.8 in \cite{book-han-kobayashi}.
\item If $\ux - \uy - \us$ form a Markov chain, then
\[
\begin{split}
\hat{S}(\ux,\uy) = & \ \expect{\us|\ux,\uy} \\
		   = & \ \expect{\us|\uy} \\
		   = & \ \hat{\us}(\uy).
\end{split}
\]
To prove the converse, suppose $\hat{S}(\ux,\uy) = \hat{\us}(\uy)$. Now let $\underline{E}$ be the error in estimation of $\us$ given $(\ux,\uy)$, which is independent of $\ux$ and $\uy$. Then,
\[
\begin{split}
\prob{}_{\us|\ux,\uy}(\underline{s}|\ux = \underline{x},\uy = \underline{y}) = & \ \prob{}_{\underline{E}}(\underline{s} - \hat{\us}(\underline{x},\underline{y})) \\
= & \ \prob{}_{\underline{E}}(\underline{s} - \hat{\us}(\underline{y})) \\
= & \ \prob{}_{\us|\uy}(\underline{s}|\uy = \underline{y}). 
\end{split}
\]
\item Observe that
\[
\begin{split}
\hat{S}(X,Y) = & \ \expect{S|X,Y} = \expect{S|X,Z} \\
		    = & \ X + \expect{N|Z} \\
		    = & \ X + \frac{\expect{NZ}}{\expect{Z^{2}}}Z \\
		   = & \ \frac{\expect{NZ}}{\expect{Z^2}}Y + \left(1 - \frac{\expect{NZ}}{\expect{Z^2}}\right)X.
\end{split}
\]
From Claim 2, it follows that $X -Y - S)$ form a Markov chain iff $\expect{NZ}  = \expect{Z^2}$.
\end{enumerate}
\end{proof}
\begin{lemma}
For any Gaussian random variables $X$, $Y$, $S_1$ and $S_2$, $I(X;\underline{S}|Y) = 0$ iff $I(X;S_1|Y)~=~0$ and $I(X;S_2|Y) = 0$.
\label{lemma-markov-chain2}
\end{lemma}
\begin{proof}
Since $I(X;S_i|Y) < I(X;\underline{S}|Y)$ for $i = 1, 2$, the `only if' part of the Lemma is clear. It remains to prove the `if' part of the Lemma and using Lemma~\ref{lemma-markov-chain1}, it is enough to show that $X -Y - \underline{S}$ form a Markov chain if $X -Y - S_1$ and $X -Y - S_2$ form Markov chains.

Let $\hat{X}(Y)$ be the MMSE estimate of $X$ given $Y$ and $E$ be the error in estimate. For $i = 1,2$, since $X - Y - S_i$ form a Markov chain, it follows from Lemma~\ref{lemma-markov-chain1} that $\hat{X}(Y,S_i) = \hat{X}(Y)$. Hence $E$ is independent of both $S_1$ and $S_2$. Since $E, S_1$ and $S_2$ are all Gaussian, $E$ is also independent of $\underline{S}$. Therefore $\hat{X}(Y,\underline{S}) = \hat{X}(Y)$ and hence $X -Y - \underline{S}$ form a Markov chain.
\end{proof}
\section{Two User Interference Channel: Existing Bounds}
\label{Sec:twouser-previouswork}
The information-theoretic study of interference channels has mainly been limited to the two-user case, with the hope that the insights obtained from studying the two-user case can be generalized to an interference network with more than two users. With $M = 2$ in \eqref{interference_network}, we get the two-user Gaussian interference channel parameterized by $\{P_1,P_2,h_{12},h_{21}\}$:
\begin{equation}
\begin{split}
Y_1 = & \ X_1 + h_{12}X_2 + Z_1 \\
Y_2 = & \ h_{21}X_1 + X_2 + Z_2
\label{two_user_ic}
\end{split}
\end{equation}
with average transmit power constraints $P_1$ and $P_2$ on users $1$ and $2$, respectively. The capacity region of this channel is known only in the very strong interference \cite{Carleial1975} and strong interference \cite{HK1981}, \cite{Sato1981} settings, where it can be established that both the users can decode all the transmitted messages, and thus the capacity region is the same as that of the compound multiple access channel. In the rest of this section, we summarize the existing bounds on the capacity region of the weak interference channel, where $h_{12} < 1$ and $h_{21} < 1$.
\subsection{Inner bounds}
\textit{Simple schemes:}
In the interference free scenario, where $h_{12} = h_{21} = 0$, single-user Gaussian codebooks at the transmitters are obviously capacity-achieving. Thus, if the interference is low, a reasonable strategy is to treat interference as noise at the receivers, and employ single-user Gaussian codebooks at the transmitters to achieve the following sum rate.
\begin{prop}[Treating interference as noise]
\label{prop:simple-schemes}
The sum capacity ($\Csum$) of the two-user Gaussian interference channel \eqref{two_user_ic} is lower bounded by
\[
\Csum \geq \logfn{1 + \frac{P_1}{1+h_{12}^2P_2}} +  \logfn{1 + \frac{P_2}{1+h_{21}^2P_1}} 
\]
\end{prop}
\vspace*{10pt}
Clearly such a strategy will not work if the interference is moderate, in which case, another simple alternative is to orthogonalize the users in time or frequency. 
\medskip

\textit{Sophisticated schemes:}
Interference, unlike noise, is generated by other users and hence has a definite structure. Sophisticated schemes that exploit the interference structure could potentially perform better than the simple schemes described above. Han and Kobayashi introduced such a sophisticated scheme in \cite{HK1981}, which results in the best known achievable region for the two-user channel. And while Chong, Motani and Garg have recently simplified  the Han-Kobayashi region \cite{CMG2007}, it still remains formidable to compute.

\subsection{Outer Bounds}
The best known outer bounds to the capacity region of the two-user Gaussian interference channel are the one due to Sato, Costa and Kramer \cite{Sato1978,Costa1985,Kramer2004}, which we refer to as the {\em broadcast channel outer bound}; and the one due to Etkin, Tse, and Wang \cite{OneBit2007}, which we refer to as the {\em ETW outer bound}. In the rest of this section, we review these outer bounds. We also give a simple and more direct proof of the \BCbound, and illustrate that it is a tightened version of the {\em \ZCbound} \cite{Kramer2004,Sason2004}. We make use of this connection to tighten the ETW outer bound in Section~\ref{Sec:twouser-outerbounds-capacityregion}.

A salient feature of these outer bounds is that they are based on a genie providing side information to the receivers. Since the receivers can choose not to use the side-information, the capacity region of the genie-aided channel is an obvious outer bound on the capacity region of the interference channel. Throughout this paper, we will assume that  the side information is linear in the inputs with additive Gaussian noise that is i.i.d. in time. Thus, the side information will be Gaussian if all the inputs are Gaussian.

Some notation is required before proceeding further.
The variable $S_r$ denotes the side information given to receiver $r$, $r=1,2$. The variable $X_{tG}$ denotes the zero-mean Gaussian random variable with variance $P_t$, $t=1,2$. The variables $Y_{rG}$ and $S_{rG}$ denote the Gaussian outputs and side information at receiver $r$, respectively, that result when all the channel inputs are Gaussian, i.e., when  $X_t = X_{tG}$, for $t=1,2$. The quantities $X_{tG}^n, Y_{rG}^n$ and $ S_{rG}^n$ denote i.i.d. sequences  of the corresponding Gaussian random variables.

\subsection{Bounding Techniques}
\label{sec:bound_tech}
Consider the following possible ways of bounding the rate ($R_1$) of user $1$
\begin{itemize}
\item \textit{No Side Information:} If the receivers do not receive any side information, then $R_1$ can be bounded using Fano's equality as follows:
\begin{equation}
\begin{split}
n(R_1 - \epsilon_n) \leq & \ I(X_1^n;Y_1^n) \\
        =  & \ \entropy(Y_1^n) - \entropy(Y_1^n|X_1^n) \\
 \leq & \ n\entropy(Y_{1G}) - \entropy(h_{12}X_2^n + Z_1^n) .
\end{split}
\label{Rx1-NoSI}
\end{equation}
\item \textit{Interference Free:} Providing receiver $1$ with the knowledge of the interfering signal $X_2$ can only increase the achievable rate $R_1$, hence
\begin{equation}
\begin{split}
n(R_1 - \epsilon_n) \leq & \ I(X_1^n;Y_1^n,X_2^n) \\
                       = & \ I(X_1^n;Y_1^n|X_2^n) \\
                       = & \ \entropy(Y_1^n|X_2^n) - \entropy(Y_1^n|X_1^n,X_2^n) \\
    \stackrel{(b)}     = & \ \entropy(X_1^n+Z_1^n) - n\entropy(Y_{1G}|X_{1G},X_{2G}) \\
    \stackrel{(c)}     = & \ \entropy(h_{21}X_1^n+h_{21}Z_1^n) - n\entropy(h_{21}Y_{1G}|X_{1G},X_{2G})
\end{split}
\label{Rx1-IF}
\end{equation}
where the step (b) follows because $Y_1|X_1,X_2$ is the Gaussian noise at the receiver,  which is not a function of the input distributions. The scaling in the step (c) is done for convenience.
\item \textit{Genie-aided: }
Here a genie provides side information $S_1$ to receiver 1. As we stated earlier, we assume that  the side information is linear in the inputs with additive Gaussian noise that is i.i.d. in time. For the two-user case, we further restrict our attention to genie signals such that, conditioned on the input sequence $X_i^n$, the sequence $S_i^n$ is i.i.d. Gaussian (this holds, for example, if $S_i^n = X_i^n + W_i^n$, where $W_i^n$ is i.i.d. Gaussian). Then we can write
\begin{equation}
\begin{split}
n(R_1 - \epsilon_n) \leq & \ I(X_1^n;Y_1^n,S_1^n) \\
      =   & \ I(X_1^n;S_1^n) + I(X_1^n;Y_1^n|S_1^n) \\
      =   & \ \entropy(S_1^n) - \entropy(S_1^n|X_1^n) + \entropy(Y_1^n|S_1^n) - \entropy(Y_1^n|S_1^n,X_1^n) \\
\stackrel{(d)} = & \ \entropy(S_1^n) - n\entropy(S_{1G}|X_{1G}) + \entropy(Y_1^n|S_1^n) - \entropy(Y_1^n|S_1^n,X_1^n) \\
\stackrel{(e)} \leq & \ \entropy(S_1^n) - n\entropy(S_{1G}|X_{1G}) + n\entropy(Y_{1G}|S_{1G}) - \entropy(Y_1^n|S_1^n,X_1^n)
\end{split}
\label{Rx1-GA}
\end{equation}
where the step (d) holds because of the assumption on the genie signal, and the step (e) follows from  Lemma~\ref{lemma-concavity-conditonal-entropy}.
\end{itemize}

The term $nR_2$ can bounded in similar ways:
\begin{itemize}
\item \textit{No Side Information: }
\begin{equation}
\begin{split}
n(R_2 - \epsilon_n)   = & \ I(X_2^n;Y_2^n) \\
                 \leq & \ n\entropy(Y_{2G}) - \entropy(h_{21}X_1 + Z_2^n) . \\
\end{split}
\label{Rx2-NoSI}
\end{equation}
\item \textit{Interference Free: }
\begin{equation}
\begin{split}
n(R_2 - \epsilon_n) \leq & \ I(X_2^n;Y_2^n|X_1^n) \\
     \leq & \ \entropy(h_{12}X_2^n+h_{12}Z_2^n) - n\entropy(h_{12}Y_{2G}|X_{1G},X_{2G}) .\\
\end{split}
\label{Rx2-IF}
\end{equation}
\item \textit{Genie-aided: }
\begin{equation}
\begin{split}
n(R_2 - \epsilon_n) \leq & \ I(X_2^n;Y_2^n,S_2^n) \\
     \leq & \ \entropy(S_2^n) - n\entropy(S_{2G}|X_{2G}) + n\entropy(Y_{2G}|S_{2G}) - \entropy(Y_2^n|S_2^n,X_2^n) .
\end{split}
\label{Rx2-GA}
\end{equation}
\end{itemize}

\subsection{Etkin, Tse and Wang (ETW) Outer Bound \cite{OneBit2007}} \label{sec:ETW}
If the genie signals are defined as
\begin{equation}
\begin{split}
S_1 = & \ h_{21}X_1 + Z_2 \\
S_2 = & \ h_{12}X_2 + Z_1
\end{split}
\label{genie_etw}
\end{equation}
then  the following relations hold true
\begin{equation}
\begin{split}
\entropy(Y_2^n|S_2^n,X_2^n) = & \ \entropy(S_1^n) \\
\entropy(Y_1^n|S_1^n,X_1^n) = & \ \entropy(S_2^n) .
\end{split}
\label{genie-relations-etw}
\end{equation}
Since $h_{12} \leq 1, h_{21} \leq 1$, we can use the worst case noise result (Lemma~\ref{lemma-worstcase-noise-scalar}) to obtain the following inequalities:
\begin{equation}
\begin{split}
\entropy(h_{21}X_1^n+h_{21}Z_1^n) - \entropy(h_{21}X_1^n+Z_2^n) \leq & \ n\entropy(h_{21}X_{1G}+h_{21}Z_1) - n\entropy(h_{21}X_{1G}+Z_2) \\
\entropy(h_{12}X_2^n+h_{12}Z_2^n) - \entropy(h_{12}X_2^n+Z_1^n) \leq & \ n\entropy(h_{12}X_{2G}+h_{12}Z_2) - n\entropy(h_{12}X_{2G}+Z_1) .\\
\end{split}
\label{worst-case-noise-zchannel}
\end{equation}

The relations \eqref{genie-relations-etw} and \eqref{worst-case-noise-zchannel}, together with bounding techniques described in the previous subsection, lead succinctly to the outer bound on the capacity region given by Etkin, Tse and Wang \cite{OneBit2007}:
\begin{lemma}[Etkin, Tse and Wang \cite{OneBit2007}]
The capacity region of a two-user Gaussian interference channel with $h_{12} \leq 1$ and $h_{21} \leq 1$ is contained in the region:
\begin{eqnarray}
R_1 \leq & \ I(X_{1G};Y_{1G}|X_{2G}) \label{R1-bound-etw}\\
R_2 \leq & \ I(X_{2G};Y_{2G}|X_{1G}) \label{R2-bound-etw}\\
R_1 + R_2 \leq & \ I(X_{1G};Y_{1G}|X_{2G}) + I(X_{2G};Y_{2G}) \label{inv-zchannel-bound-etw} \\
R_1 + R_2 \leq & \ I(X_{1G};Y_{1G}) + I(X_{2G};Y_{2G}|X_{1G}) \label{zchannel-bound-etw}\\
R_1 + R_2 \leq & \ I(X_{1G};Y_{1G},S_{1G}) + I(X_{2G};Y_{2G},S_{2G}) \label{onebit-bound-etw}\\
2R_1 + R_2 \leq & \ I(X_{1G};Y_{1G}|X_{2G}) + I(X_{1G};Y_{1G}) + I(X_{2G};Y_{2G},S_{2G}) \label{2R1-R2-bound-etw}\\
R_1 + 2R_2 \leq & \ I(X_{1G};Y_{1G},S_{1G}) + I(X_{2G};Y_{2G}|X_{1G}) + I(X_{2G};Y_{2G}) \label{R1-2R2-bound-etw}
\end{eqnarray}
where the genie signals $\{S_1,S_2\}$ are defined in (\ref{genie_etw}).
\label{outerbounds-etw}
\end{lemma}
\begin{proof}
The outer bounds immediately follow by choosing the appropriate bounding technique from Section~\ref{sec:bound_tech},  and using the relations \eqref{worst-case-noise-zchannel} and \eqref{genie-relations-etw}, where necessary. For example, to derive the bound \eqref{zchannel-bound-etw}, use \eqref{Rx1-NoSI} and \eqref{Rx2-IF} and use the worst case noise result \eqref{worst-case-noise-zchannel} to show that $\entropy(h_{12}X_2^n+h_{12}Z_2^n) - \entropy(h_{12}X_2^n+Z_1^n)$ is maximized by $X_{2G}^n$. To derive the bound \eqref{onebit-bound-etw}, use \eqref{Rx1-GA} and \eqref{Rx2-GA} and use \eqref{genie-relations-etw} to show that the right hand side (RHS) of \eqref{Rx1-GA}\ plus the RHS of \eqref{Rx2-GA} is maximized by $X_{1G}^n$ and $X_{2G}^n$.
\end{proof}
\begin{remark}
The RHS terms in the outer bounds can easily be shown to be equivalent to those in Theorem~3 of \cite{OneBit2007}
by making the following substitutions:
\[
\begin{split}
I(X_{1G};Y_{1G}|X_{2G}) = & \ \logfn{1+P_1} \\
I(X_{1G};Y_{1G}) = & \ \logfn{1+\frac{P_1}{1 + h_{12}^2P_2}} \\
I(X_{1G};Y_{1G},S_{1G}) = & \ \logfn{1 + h_{21}^2P_1+ \frac{P_1}{1 + h_{12}^2P_2}}
\end{split}
\]
and similar substitutions for the terms corresponding to  user $2$.
\end{remark}

The form of the outer bound given in Lemma~\ref{outerbounds-etw} is strikingly similar to the simplified HK region  \cite{CMG2007}, and in fact a special case of the HK region is shown to be {\em within one bit} of the outer bound \cite{OneBit2007},\cite{TelatarTse2007}.
\subsection{Outer Bounds to One-Sided Interference Channels}
In deriving the bounds \eqref{inv-zchannel-bound-etw} and \eqref{zchannel-bound-etw}, one of the receivers is made interference free. Thus these outer bounds are derived for the one-sided interference channel, where only one user experiences the interference. Such a channel is also called the Z-channel, and  we therefore  refer to the outer bounds \ \eqref{inv-zchannel-bound-etw} and \eqref{zchannel-bound-etw} as the \ZCbound{}s.

In \cite{Costa1985}, Costa showed the equivalence between the Z-channel and the degraded interference channel, and in \cite{Sato1978}, Sato showed that the capacity region of the degraded interference channel is contained in the capacity region of a broadcast channel. Using these ideas, Kramer established an outer bound to the capacity region of the Z-channel \cite{Kramer2004}. We refer to this outer bound as the \BCbound. We show that \BCbound{} is a tightened version of \ZCbound{}, and thus provide a simple and direct proof of the \BCbound. In deriving the Z-channel sum rate outer bound \eqref{zchannel-bound-etw}, we have used the worst case noise result to relate the terms $\entropy(h_{12}X_2^n+h_{12}Z_2^n)$ and $\entropy(h_{12}X_2^n+Z_1^n)$. Instead, the  EPI can be used to obtain a tighter relation, which results in the $\BCbound$.
\begin{lemma}[Broadcast channel outer bound \cite{Sato1978,Costa1985,Kramer2004}]
The capacity region of a two-user Gaussian interference channel with $h_{12} \leq 1$ and $h_{21} \leq 1$ is contained in the region
\begin{equation}
R_1 \leq \logfn{\frac{1 + P_1 + h_{12}^2P_2}{1 + h_{12}^2(2^{2R_2}-1)}} .
\end{equation}\label{lemma:bcb}
\end{lemma}
By changing the order of the users, we also have
\begin{equation}
R_2 \leq \logfn{\frac{1 + P_2 + h_{21}^2P_1}{1 + h_{21}^2(2^{2R_1}-1)}} .
\end{equation}
\begin{proof}
Using (\ref{Rx1-NoSI}) and (\ref{Rx2-IF}), we have
\[
\begin{split}
nR_1 \leq & \ \frac{n}{2}\log(2\pi{}e(1+P_1+h_{12}^2P_{2})) - \entropy(h_{12}X_2^n + Z_1^n) \\
nR_2 \leq & \ \entropy(h_{12}X_2^n + h_{12}Z_2^n) - \frac{n}{2}\log(2\pi{}eh_{12}^2) .
\end{split}
\]
From the EPI (Corollary~\ref{corollary-EPI}), it follows that
\[
\begin{split}
\entropy(h_{12}X_2^n + Z_1^n) \geq & \ \frac{n}{2}\log\left(2\pi{}e(1-h_{12}^2) + 2^{\frac{2}{n}\entropy(h_{12}X_2^n + h_{12}Z_2^n)}\right)\\
\geq & \ \frac{n}{2}\log\left(2\pi{}e(1-h_{12}^2) + 2\pi{}eh_{12}^2 2^{2R_2}\right) .
\end{split}
\]
Therefore,
\[
\begin{split}
nR_1 \leq & \ \frac{n}{2}\log(2\pi{}e(1+P_1+h_{12}^2P_2)) - \frac{n}{2}\log\left(2\pi{}e(1-h_{12}^2) + 2\pi{}eh_{12}^2 2^{2R_2}\right) \\
= & \ \frac{n}{2}\log\left(\frac{1 + P_1 + h_{12}^2P_2}{1 - h_{12}^2 + h_{12}^2 2^{2R_2} }\right) .
\end{split}
\]
\end{proof}
\begin{remark}
The outer bound in Lemma~\ref{lemma:bcb} can be shown to be identical to that presented in Theorem~2 of \cite{Kramer2004}.
\end{remark}
\subsection{Tightening the Outer Bounds}
\label{sec:obersvations}
The outer bounds presented above  can be tightened by using the following observations:
\begin{itemize}
\item \textit{Generalized Genie: }In \cite{OneBit2007}, the genie \eqref{genie_etw} is selected to satisfy (\ref{genie-relations-etw}). However the techniques developed in \cite{OneBit2007} can  be generalized to a larger class of genie signals, by using the worst case noise result to relate the terms in  (\ref{genie-relations-etw}) instead of canceling the terms. In fact one of the main results of this paper,  the sum capacity of the two-user Gaussian interference channel in the low interference regime, is a direct consequence of this observation.
\item \textit{EPI-based bounds: }We have shown that the \BCbound~is a tightened version of sum rate bounds of  (\ref{inv-zchannel-bound-etw}) and (\ref{zchannel-bound-etw})  by using the  EPI instead of the worst case noise result. We can similarly apply the EPI to the other outer bounds in the Lemma~\ref{outerbounds-etw}.
\end{itemize}

We now proceed to use these observations to tighten the existing outer bounds.
\section{Two User Interference Channel: Sum Capacity in Low Interference Regime}
\label{Sec:twouser-sumcapacity}
%
Consider  the limiting scenario where the interference parameters $\{h_{rt}\}_{r \neq t}$ go to zero uniformly. In the limit, when there is no interference, single user Gaussian codes are optimal. Given this fact, a natural question to ask is the following: {\em In terms of the optimality of single user Gaussian codes,  is the transition from ``no interference" to ``interference" continuous?} If any other strategy performs better than treating interference as noise, then this implies that the receivers are able to exploit the structure in the interference. On the other hand, for low enough interference levels the receivers may not be able to exploit such  structure. Thus it is reasonable to expect the transition to be continuous. In this section, we establish this notion mathematically by showing that treating interference as noise indeed achieves the sum capacity in a low (but nonzero) interference regime.
\subsection{Symmetric Interference Channel}
The essential ideas and results on the sum capacity of the two-user interference channel are captured in the {\em symmetric }interference channel, for which  $P_1 = P_2 = P$ and $h_{12} = h_{21} = h$. For this channel we shall establish the following result.
\begin{theorem}
For the symmetric interference channel, if the interference parameter $h$ satisfies the condition
\begin{eqnarray}
|h +h^3P| \leq 0.5
\label{condition_mainresult}
\end{eqnarray}
then treating interference as noise achieves the sum capacity, which is given by
\begin{equation} \label{eq:Csumsymm}
\Csum = \log\left(1+\frac{P}{1+h^2P}\right).
\end{equation}
\label{mainresult}
\end{theorem}
Since the achievability part of the theorem is obvious, we only need to establish an upper bound on $\Csum$ that matches the expression given on the RHS of \eqref{eq:Csumsymm}.  We use the concept of the genie-aided outer bound (see Section~\ref{sec:bound_tech}), but with a class of genie signals that is more general than that used for the ETW bound of Section~\ref{sec:ETW}. In particular,  we wish to choose the genie to produce the tightest possible upper bound. To this end, we introduce the following two qualities of a good genie.
\begin{itemize}
\item \textit{Useful Genie: }The ETW genie (\ref{genie_etw}) is useful in deriving an outer bound on the sum capacity of the interference channel. The reason behind its usefulness is the property (\ref{genie-relations-etw}) that facilitates the derivation of the sum capacity of the genie-aided channel. Using (\ref{genie-relations-etw}), it can be shown that Gaussian inputs, which  are i.i.d. in time and satisfy the power constraint with equality,  are capacity achieving for the genie-aided channel. Hence the sum capacity of the genie-aided channel equals
    \begin{equation}
    I(X_{1G};Y_{1G},S_{1G}) + I(X_{2G};Y_{2G},S_{2G}) .
    \label{useful-genie-sumcapacity}
    \end{equation}
Interestingly, there exists a larger class of genie signals for which the optimality of Gaussian inputs holds. We therefore define  a genie to be {\em useful}, if it results in a genie-aided channel whose sum capacity (is achieved by Gaussian inputs and) is  given by (\ref{useful-genie-sumcapacity}).

A second example of a useful genie signal is the interference removal genie, i.e., the genie that provides side information $S_1 = X_2$ to receiver 1 and side information $S_2 = X_1$ to receiver 2. Such a genie is clearly useful because the resulting genie-aided channel is the parallel Gaussian channel whose sum capacity is easily seen to be given by (\ref{useful-genie-sumcapacity}). However, being too generous, such a genie does not result in a tight upper bound.
    This leads us to the notion of a {\em smart} genie.
\item \textit{Smart Genie: }A smart genie results in a tight upper bound on the sum capacity. More precisely, if Gaussian inputs are used, then the presence of the genie does not improve the sum rate, i.e.,
    \[
    \begin{split}
    I(X_{1G};Y_{1G},S_{1G}) = & \ I(X_{1G};Y_{1G}) \\
    I(X_{2G};Y_{2G},S_{2G}) = & \ I(X_{2G};Y_{2G}) .
    \end{split}
    \]
    An example of the smart genie is one that does not interact with the receivers at all; however, it is obviously not useful.
\end{itemize}
If the genie is useful and smart, then the sum capacity is upper bounded by $I(X_{1G};Y_{1G},S_{1G}) + I(X_{2G};Y_{2G},S_{2G}) = I(X_{1G};Y_{1G}) + I(X_{2G};Y_{2G})$, which is the sum rate achieved by treating interference as noise. Thus it is enough to show the existence of a genie that is both useful and smart to prove Theorem~\ref{mainresult}. So the essential question is: \emph{Is there a ``divine'' genie that is both useful and smart?}

\begin{figure}[!t]
\centering
\includegraphics[width=0.5\textwidth]{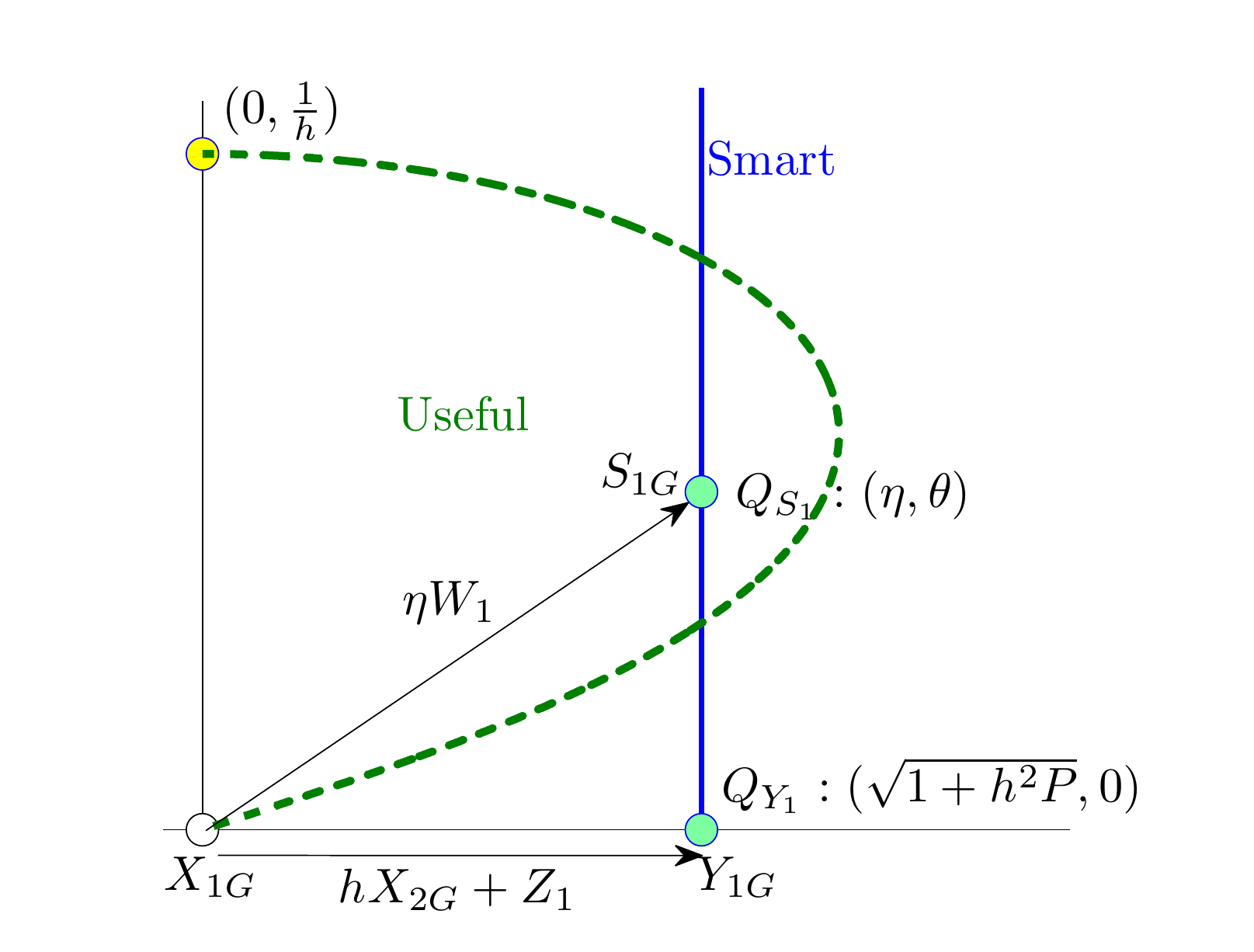}
\caption{The figure is a Hilbert space representation of the channel input, channel output and genie signal. The genie is a) useful if it lies inside the dashed curve,  and b) smart if it lies on the solid line. If the dashed curve and solid line intersect, treating interference as noise achieves sum capacity.}
\label{polar}
\end{figure}

\begin{figure}[!t]
\centering
\includegraphics[width=0.7\textwidth]{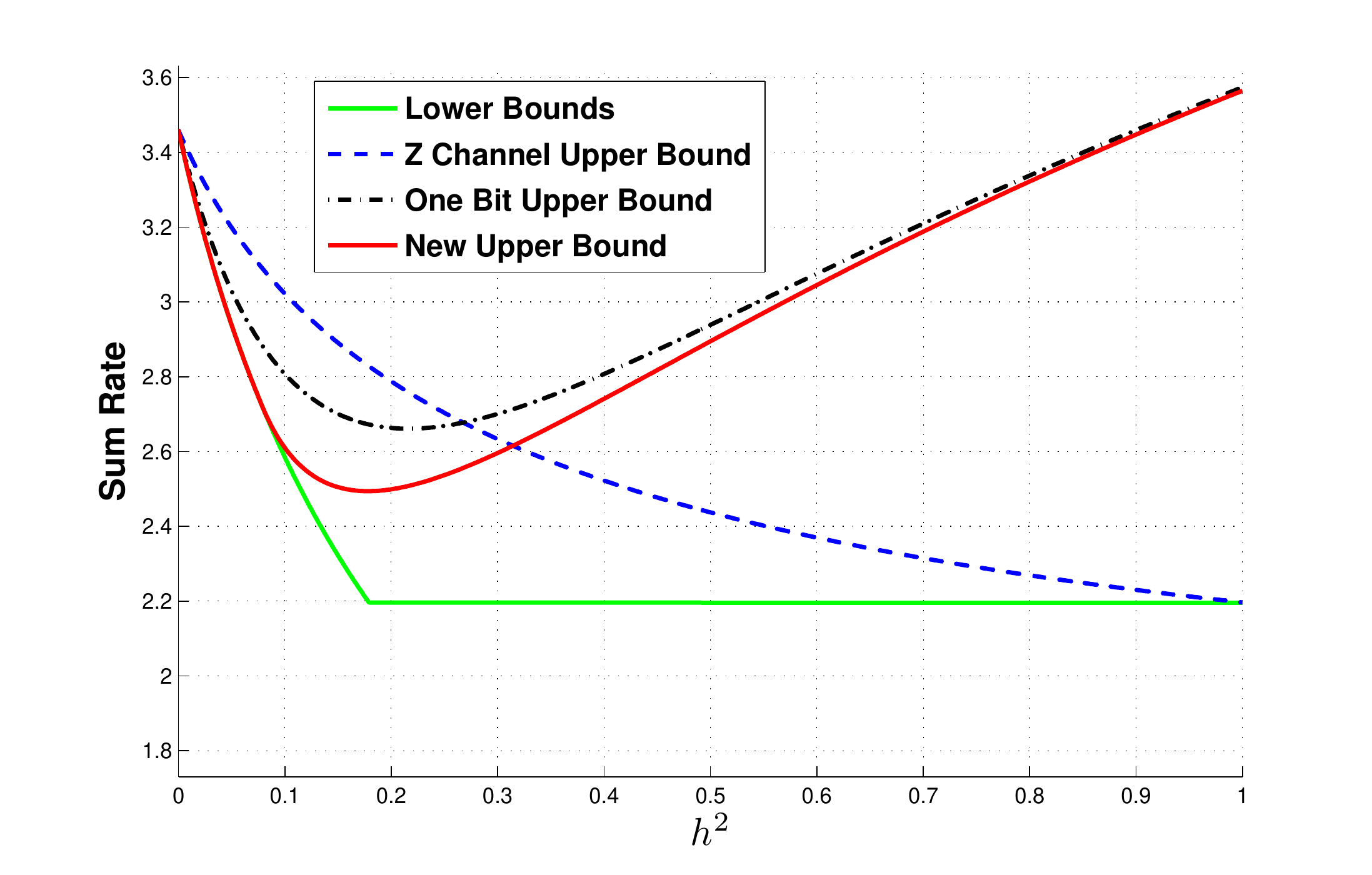}
\caption{Sum capacity of the two-user symmetric Gaussian interference channel in the low interference regime.}
\label{fig:sumcapacity_twouser}
\end{figure}

\begin{figure}[!t]
\centering
\includegraphics[width=0.7\textwidth]{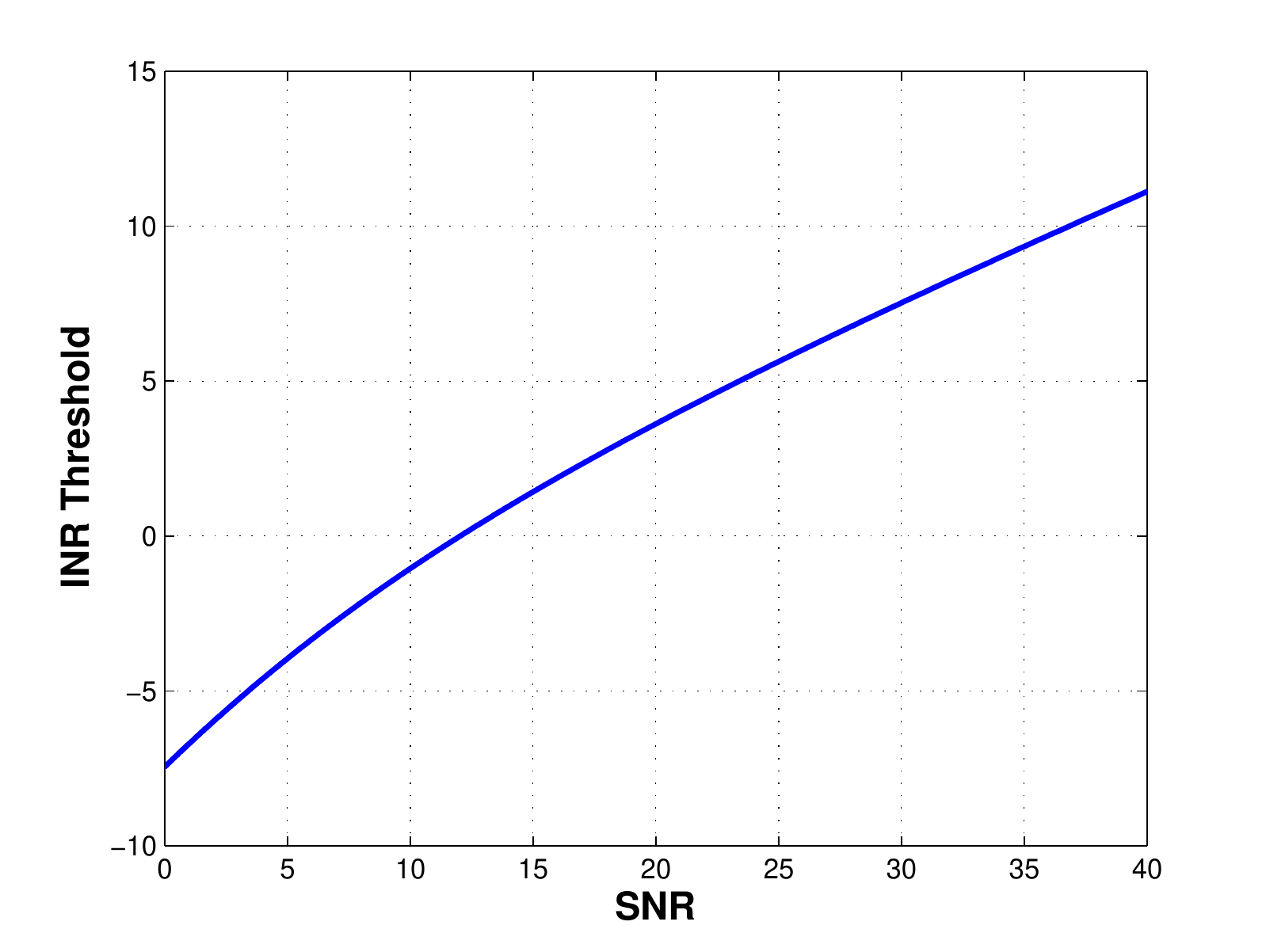}
\caption{Two user symmetric Gaussian interference channel: INR threshold, below which treating interference as noise achieves the sum capacity, as a function of the SNR.}
\label{fig:SNRvsINR}
\end{figure}

The quest for the divine genie can be simplified by imposing a structure on the side information it provides. Following \eqref{genie_etw}, we set:
\begin{equation} \label{genie_form}
\begin{split}
S_1 = & \ hX_1 + h\eta{}W_1 \\
S_2 = & \ hX_2 + h\eta{}W_2
\end{split}
\end{equation}
where $W_1, W_2 \sim \mathcal{N}(0,1)$ and $\eta$ is a positive real number. However, unlike in \eqref{genie_etw}, we allow  $W_1$ to be correlated to $Z_1$ (and $W_2$ with $Z_2$), with correlation coefficient  $\rho$.
\begin{lemma}[Useful Genie]
The sum capacity of the genie-aided channel with side information given in \eqref{genie_form} is achieved by using Gaussian inputs and by treating interference as noise at the receiver  if the following condition holds:
\begin{eqnarray}
|h\eta| \leq \sqrt{1-\rho^2} .
\label{condition_useful}
\end{eqnarray}
Hence the sum capacity of the symmetric interference channel  is bounded as
\begin{equation}
\Csum \leq I(X_{1G};Y_{1G},S_{1G}) + I(X_{2G};Y_{2G},S_{2G}).
\label{sumcapacity_genieaided}
\end{equation}
\label{useful_genie}
\end{lemma}
\begin{proof}
Add (\ref{Rx1-GA}) and (\ref{Rx2-GA}) to get the following outer bound on $n(R_1+R_2-2\epsilon)$.
\[
\begin{split}
& \entropy(S_1^n) - n\entropy(S_{1G}|X_{1G})  + n\entropy(Y_{1G}|S_{1G}) - \entropy(Y_1^n|S_1^n,X_1^n)  \\
& + \entropy(S_2^n)  - n\entropy(S_{2G}|X_{2G}) + n\entropy(Y_{2G}|S_{2G}) - \entropy(Y_2^n|S_2^n,X_2^n) .
\end{split}
\]
Thus it only remains to show that
\[
\entropy(S_1^n) -  \entropy(Y_2^n|S_2^n,X_2^n) + \entropy(S_2^n) -  \entropy(Y_1^n|S_1^n,X_1^n)
\]
is maximized by $X_{1G}^n$ and $X_{2G}^n$. Now consider
\[
\begin{split}
\entropy(S_1^n) -  \entropy(Y_2^n|S_2^n,X_2^n) = & \ \entropy(hX_1^n+h\eta{}W_1^n) - \entropy(hX_1^n+Z_2^n|W_2^n)\\
\stackrel{(a)} = & \ \entropy(hX_1^n+h\eta{}W_1^n) - \entropy(hX_1^n+V^n) \\
\stackrel{(b)} \leq & \ n\entropy(hX_{1G}+h\eta{}W_1) - n\entropy(hX_{1G}+V)
\end{split}
\]
where $V \sim \mathcal{N}(0,1-\rho^2)$, independent of $X_1$. Step (a) follows from Lemma~\ref{lemma-conditonal-entropy-mmse} and step (b) follows form condition (\ref{condition_useful}) and the worst case noise Lemma~\ref{lemma-worstcase-noise-scalar}. Thus $\entropy(S_1^n) -  \entropy(Y_2^n|S_2^n,X_2^n)$ is maximized by $X_{1G}^n$ and similarly $\entropy(S_2^n) -  \entropy(Y_1^n|S_1^n,X_1^n)$ is maximized by $X_{2G}^n$.
\end{proof}
\smallskip
\begin{lemma}[Smart Genie]
If Gaussian inputs are used, the interference is treated as noise, and the following condition holds
\begin{equation}
\eta\rho = 1+h^2P
\label{condition_smart}
\end{equation}
then the genie does not increase the achievable sum rate, i.e.,
\begin{equation} \label{smart_genie_definition}
\begin{split}
I(X_{1G};Y_{1G}, S_{1G}) = & \ I(X_{1G};Y_{1G}) \\
I(X_{2G};Y_{2G}, S_{2G}) = & \ I(X_{2G};Y_{2G}) .
\end{split}
\end{equation}
The converse is also true, i.e., \eqref{smart_genie_definition} implies \eqref{condition_smart}
\label{smart_genie}
\end{lemma}
\begin{proof} Since
\[
I(X_{iG};Y_{iG}, S_{iG}) = I(X_{iG};Y_{iG}) + I(X_{iG};S_{iG}|Y_{iG})
\]
\eqref{smart_genie_definition} is equivalent to
\[
\begin{split}
I(X_{iG};S_{iG}|Y_{iG}) = & \ 0 \\
\iff I(X_{iG};X_{iG}+\eta{}W_{i}|X_{iG} + hX_{jG} + Z_{1}) = & \ 0 \\
\iff \expect{\eta{}W_{i}(hX_{jG}+Z_{i})} \stackrel{(a)} = & \ \expect{(hX_{jG}+Z_{i})^{2}} \\
\iff \eta{}\rho = & \ 1 + h^{2}P.
\end{split}
\]
where the step (a) follows from Lemma~\ref{lemma-markov-chain1} and the index $j = 2$ if $i = 1$ and vice versa.
\end{proof}
\smallskip
In Figure~\ref{polar}, we plot  the usefulness and smartness constraints \eqref{condition_useful} and \eqref{condition_smart} in the Hilbert space $L^{2}$ of random variables.  Figure~\ref{polar} only shows the plane containing the transmitted signal $X_{1G}$, the received signal  $Y_{1G} = X_{1G} + hX_{2G} + Z_{1} $ and the genie signal $\frac{S_{1G}}{h} =  X_{1G} + \eta{}W_{1}$ with origin shifted to $X_{1G}$. We can view the usefulness and smartness constraints \eqref{condition_useful} and \eqref{condition_smart} on the genie as regions in the $L^{2}$ space:
\begin{itemize}
\item {\em Useful Genie: } The genie is useful, if it lies inside the dashed curve in Fig.~\ref{polar}. The boundary of the curve is obtained using the usefulness condition \eqref{condition_useful}.
\item {\em Smart Genie: } The genie is smart, if it lies on the solid line in Fig.~\ref{polar}. This is expected because $X_{1G}  -  (X_{1G} + hX_{2G} + Z_{1} ) - (X_{1G} + \eta{}W_{1})$ form a Markov chain iff $X_{1G} + \eta{}W_{1}$ is a degraded version of $X_{1G} + hX_{2G} + Z_{1}$.
\end{itemize}
There exists a genie that is both useful and smart if the usefulness region intersects with the smartness line in Fig.~\ref{polar}, i.e., if there exist $\eta$ and $\rho$ satisfying the conditions of both Lemma \ref{useful_genie} and Lemma~\ref{smart_genie}. Eliminating $\eta$ from \eqref{condition_useful} and \eqref{condition_smart} we get
\[
|h+h^3P| \leq |\rho|\sqrt{1-\rho^2}
\]
which is possible  iff
\[
|h+h^3 P| \leq 0.5 .
\]
This completes the proof of Theorem~\ref{mainresult}.  \endproof
\begin{remark}
Lemma~\ref{useful_genie} is valid even if the interference channel is not in the low interference regime. Therefore, minimizing the expression \eqref{sumcapacity_genieaided} over all possible genie signals satisfying the usefulness constraint \eqref{condition_useful} results in a valid outer bound. The notion of the smart genie, therefore, can be thought of as an intuitive way of identifying the genie that minimizes \eqref{sumcapacity_genieaided}. In \cite{ITA2008}, we use the geometric interpretation of the Figure~\ref{polar} to identify the useful genie that minimizes \eqref{sumcapacity_genieaided} when the channel is not in the low interference regime.
\end{remark}
In Figure~\ref{fig:sumcapacity_twouser}, we plot the new outer bound along with the  \ZCbound{} \eqref{zchannel-bound-etw} and the ETW outer bound \eqref{onebit-bound-etw}. Observe that the new outer bound matches with the inner bound obtained by treating interference as noise when the interference is below a treshold. 
Figure~\ref{fig:SNRvsINR} shows the interference to noise ratio (INR) threshold, below which treating interference as noise achieves the sum capacity, as a function of the signal to noise ratio (SNR) in dB scale. It can be easily shown that the INR threshold, in a dB scale, is equal to one third of the SNR in the high SNR asymptotic regime.

\subsection{Asymmetric Interference Channel}
For the asymmetric interference channel, we consider the asymmetric genie:
\begin{equation}
\begin{split}
S_1 = & \ h_{21}(X_1 + \eta_1W_1) \\
S_2 = & \ h_{12}(X_2 + \eta_2W_2) .
\end{split}
\label{genie_form_asym}
\end{equation}
Let $\rho_1$ be the correlation between $Z_1$ and $W_1$ (and $\rho_2$ the correlation between $Z_2$ and $W_2$).
\begin{theorem}
Consider  the asymmetric interference channel with interference parameters $h_{12}$ and $h_{21}$ satisfying
\begin{equation}
|h_{12}(1+h_{21}^2P_1)| + |h_{21}(1+h_{12}^2P_2)| \leq 1 .
\label{condition_asym_simple}
\end{equation}
Then treating interference as noise achieves sum capacity, which is given by
\[
\Csum = \frac{1}{2}\log\left(1+\frac{P_1}{1+h_{12}^2P_2}\right)  + \frac{1}{2}\log\left(1+\frac{P_2}{1+h_{21}^2P_1}\right).
\]
\label{sum_capacity_asym}
\end{theorem}
\begin{proof}
The proof is similar to that for the symmetric interference channel. Using the same arguments as in Lemma~\ref{useful_genie}, the genie is useful if
\[
\begin{split}
|h_{21}\eta_1| \leq & \ \sqrt{1-\rho_2^2} \\
|h_{12}\eta_2| \leq & \ \sqrt{1-\rho_1^2} .
\end{split}
\]
Also, as in Lemma \ref{smart_genie}, the genie is smart iff
\[
\begin{split}
\eta_1\rho_1 = & \ 1+h_{12}^2P_2 \\
\eta_2\rho_2 = & \ 1+h_{21}^2P_1 .
\end{split}
\]
Thus there exists a useful and smart genie if there exist  $\rho_1 \in [0,1]$ and $\rho_2 \in [0,1]$ such that
\begin{equation}
\begin{split}
|h_{12}(1+h_{21}^2P_1)| & \leq  \rho_2\sqrt{1-\rho_1^2} \\
|h_{21}(1+h_{12}^2P_2)| & \leq  \rho_1\sqrt{1-\rho_2^2} .
\label{condition_asym}
\end{split}
\end{equation}
By setting $\rho_1 = \cos\phi_1$ and $\rho_2 = \cos\phi_2$, \eqref{condition_asym} implies \eqref{condition_asym_simple}. It is also true that \eqref{condition_asym_simple} implies \eqref{condition_asym}. This can be seen by setting $\phi$ such that
 \[
 |h_{12}(1+h_{21}^2P_1)| \leq cos^2\phi \leq 1 - |h_{21}(1+h_{12}^2P_2)| .
 \]
 i.e.,
 \[
 \begin{split}
 |h_{12}(1+h_{21}^2P_1)| \leq & \cos^2\phi \\
 |h_{21}(1+h_{12}^2P_1)| \leq & \sin^2\phi .
 \end{split}
 \]
 Setting $\rho_1 = \sin\phi$ and $\rho_2 = \cos\phi$, we have \eqref{condition_asym}.
\end{proof}
\begin{remark}
Theorem~\ref{sum_capacity_asym} is establised independently in \cite{Shang-Kramer-Chen-2007} and \cite{Motahari-Khandani-2007}.
\end{remark}

\section{Two-User Interference Channel: Outer Bounds to the Capacity Region}
\label{Sec:twouser-outerbounds-capacityregion}
In Section~\ref{sec:obersvations}, we observed that the ETW outer bound in the Lemma~\ref{outerbounds-etw} can be tightened by considering a general class of genie signals and using the EPI instead of the worst case noise result. In this section, we use these observations to improve the outer bounds. 

\begin{theorem}[EPI-Based ETW Outer Bound]
\label{Th:Epied-Bound}
The capacity region of a two-user Gaussian interference channel with $h_{12} \leq 1$ and $h_{21} \leq 1$ is outer bounded by the regions given below in  Lemmas~\ref{lemma:Epied-OneBit-Bound} and \ref{lemma:Epied-2R1R2-Bound}, along with the Lemma~\ref{lemma:bcb}.
\end{theorem}
\begin{lemma}[Tightened version of the outer bound on $R_1 + R_2$ \eqref{onebit-bound-etw}] \label{lemma:Epied-OneBit-Bound}
The capacity region of a two-user Gaussian interference channel with $h_{12} \leq 1$ and $h_{21} \leq 1$ is contained in the region
\[
\begin{split}
R_2 \leq & \ \logfn{\frac{\cov{Y_{2G}|S_{2G}}}{\cov{S_{2G}|X_{2G}}}} + \logfn{\frac{\frac{\cov{S_{1G}}\cov{Y_{1G}|S_{1G}}}{\cov{S_{1G}|X_{1G}}}2^{-2R_1} - \sigma_1^2}{\cov{S_{1G}} + \sigma_2^2 } } .
\end{split}
\]
for all $\{\eta_1,\eta_2,\rho_1,\rho_2\}$, the parameters of the genie defined in (\ref{genie_form_asym}), such that
\[
\begin{split}
\sigma_1^2 = & \ 1-\rho_2^2 - (h_{21}\eta_1)^2 \ > \ 0\\
\sigma_2^2 = & \ 1-\rho_1^2 - (h_{12}\eta_2)^2 \ > \ 0 .
\end{split}
\]
Interchanging the user indices, we get another such bound.
\end{lemma}
\begin{proof}
Using (\ref{Rx1-GA}) and (\ref{Rx2-GA}), we have
\begin{equation}
\begin{split}
nR_1 \leq & \ \entropy(S_1^n) - n\entropy(S_{1G}|X_{1G}) + n\entropy(Y_{1G}|S_{1G}) - \entropy(Y_1^n|S_1^n,X_1^n) \\
nR_2 \leq & \ \entropy(S_2^n) - n\entropy(S_{2G}|X_{2G}) + n\entropy(Y_{2G}|S_{2G}) - \entropy(Y_2^n|S_2^n,X_2^n) .
\end{split}
\end{equation}
Denote
\begin{equation}
\begin{split}
r_1 = R_1 + \entropy(S_{1G}|X_{1G}) - \entropy(Y_{1G}|S_{1G}) = R_1 - \logfn{\frac{\cov{Y_{1G}|S_{1G}}}{\cov{S_{1G}|X_{1G}}}}\\
r_2 = R_2 + \entropy(S_{2G}|X_{2G}) - \entropy(Y_{2G}|S_{2G}) = R_2 - \logfn{\frac{\cov{Y_{2G}|S_{2G}}}{\cov{S_{2G}|X_{2G}}}} \\
\end{split}
\end{equation}
to obtain
\begin{equation}
\begin{split}
nr_1 \leq & \ \entropy(S_1^n) - \entropy(Y_1^n|S_1^n,X_1^n) \\
nr_2 \leq & \ \entropy(S_2^n) - \entropy(Y_2^n|S_2^n,X_2^n) .
\end{split}
\end{equation}
To apply EPI, $\{\eta_1,\eta_2,\rho_1,\rho_2\}$ should satisfy
\[
\begin{split}
(h_{21}\eta_1)^2 & \leq 1-\rho_2^2 \\
(h_{12}\eta_2)^2 & \leq 1-\rho_1^2 .
\end{split}
\]
Define the slack variables
\[
\begin{split}
\sigma_1^2 = & \ 1-\rho_2^2 - (h_{21}\eta_1)^2\\
\sigma_2^2 = & \ 1-\rho_1^2 - (h_{12}\eta_2)^2 .
\end{split}
\]
Using EPI (Corollary~\ref{corollary-EPI}), we have
\[
\begin{split}
nr_2 \leq & \ \entropy(S_2^n) - \entropy(Y_2^n|S_2^n,X_2^n) \\
     \leq & \ n\logfn{2^{\twobyn \entropy(Y_1^n|S_1^n,X_1^n)} - 2\pi{}e\sigma_1^2 } - n\logfn{2^{\twobyn \entropy(S_1^n)} + 2\pi{}e\sigma_2^2 } \\
r_2  \leq & \ \logfn{2^{\twobyn \entropy(S_1^n)}2^{-2r_1} - 2\pi{}e\sigma_1^2 } - \logfn{2^{\twobyn \entropy(S_1^n)} + 2\pi{}e\sigma_2^2 } \\
     \leq & \ \logfn{2^{2\entropy(S_{1G})}2^{-2r_1} - 2\pi{}e\sigma_1^2 } - \logfn{2^{2\entropy(S_{1G})} + 2\pi{}e\sigma_2^2 } \\
    \leq & \ \logfn{\frac{\cov{S_{1G}}2^{-2r_1} - \sigma_1^2}{\cov{S_{1G}} + \sigma_2^2 } } .
\end{split}
\]
By eliminating $r_1$ and $r_2$, we get
\[
\begin{split}
R_2 \leq & \ \logfn{\frac{\cov{Y_{2G}|S_{2G}}}{\cov{S_{2G}|X_{2G}}}} + \logfn{\frac{\frac{\cov{S_{1G}}\cov{Y_{1G}|S_{1G}}}{\cov{S_{1G}|X_{1G}}}2^{-2R_1} - \sigma_1^2}{\cov{S_{1G}} + \sigma_2^2 } } .
\end{split}
\]
\end{proof}
\begin{remark}
Lemma~\ref{lemma:Epied-OneBit-Bound}, being a tightened version of the ETW sum rate outer bound \eqref{onebit-bound-etw}, includes the new sum rate outer bounds presented in Section~\ref{Sec:twouser-sumcapacity}.
\end{remark}
\begin{lemma}[Tightened versions of the outer bounds on $2R_1+R_2$ \eqref{2R1-R2-bound-etw} and $R_1+2R_2$ \eqref{R1-2R2-bound-etw}] \label{lemma:Epied-2R1R2-Bound}
The capacity region of a two-user Gaussian interference channel with $h_{12} \leq 1$ and $h_{21} \leq 1$ is contained in the region
\[
R_2 \leq \ \logfn{ \frac{\cov{Y_{2G}|S_{2G}}}{\cov{S_{2G}|X_{2G}}}} + \logfn{\frac{\cov{Y_{1G}}2^{-2R_1} - \sigma_1^2}{h_{21}^22^{2R_1} + \sigma_2^2} }
\]
for all $\{\eta_1,\eta_2,\rho_1,\rho_2\}$, the parameters of the genie defined in (\ref{genie_form_asym}), such that
\[
\begin{split}
\sigma_1^2 = & \ 1 - (h_{12}\eta_2)^2 \ > \ 0\\
\sigma_2^2 = & \ 1-\rho_2^2- h_{21}^2 \ > \ 0 .
\end{split}
\]
Interchanging the user indices, we get another such bound.
\end{lemma}
\begin{proof}
Use (\ref{Rx1-NoSI}), (\ref{Rx1-IF}) and (\ref{Rx2-GA}) to obtain:
\begin{equation} \label{proof-epied2R1R2-bounds}
\begin{split}
nR_1 \leq & \ n\entropy(Y_{1G}) - \entropy(h_{12}X_2 + Z_1^n) \\
nR_1 \leq & \ \entropy(h_{21}X_1^n+h_{21}Z_1^n) - n\entropy(h_{21}Y_{1G}|X_{1G},X_{2G}) \\
nR_2 \leq & \ \entropy(S_2^n) - n\entropy(S_{2G}|X_{2G}) + n\entropy(Y_{2G}|S_{2G}) - \entropy(Y_2^n|S_2^n,X_2^n)
\end{split}
\end{equation}
To apply EPI, $\{\eta_1,\eta_2,\rho_1,\rho_2\}$ should satisfy
\[
\begin{split}
h_{12}\eta_2 \leq & \ 1 \\
h_{21} \leq & \ \sqrt{1-\rho_2^2} .
\end{split}
\]
Define the slack variables
\[
\begin{split}
\sigma_1^2 = & \ 1 - (h_{12}\eta_2)^2 \\
\sigma_2^2 = & \ 1-\rho_2^2- h_{21}^2 .
\end{split}
\]
Using EPI (Corollary~\ref{corollary-EPI}) and the bounds on $nR_1$ in \eqref{proof-epied2R1R2-bounds}, we obtain
\[
\begin{split}
\entropy(S_2^n) \stackrel{(a)} \leq & \ \frac{n}{2} \log\left(2^{\twobyn{}\entropy(h_{12}X_2 + Z_1^n)} - 2\pi{}e\sigma_1^2\right)\\
								\stackrel{(b)} \leq & \ \frac{n}{2} \log\left(2^{2\entropy(Y_{1G})}2^{-2R_1} - 2\pi{}e\sigma_1^2\right)
\end{split}
\]
and
\[
\begin{split}
\entropy(Y_2^n|S_2^n,X_2^n) \stackrel{(c)} \geq & \ \frac{n}{2} \log\left(2^{\twobyn{}\entropy(h_{21}X_1 + h_{21}Z_1^n)} + 2\pi{}e\sigma_2^2\right)\\								
								\stackrel{(d)} \geq & \ \frac{n}{2} \log\left(2\pi{}e2^{2R_1}h_{21}^2 + 2\pi{}e\sigma_2^2\right)\\
\end{split}
\]
where the steps $(a)$ and $(c)$ follow from EPI (Corollary~\ref{corollary-EPI}) and the steps $(b)$ and $(d)$ use the bounds on $nR_1$ in \eqref{proof-epied2R1R2-bounds}. Using the above relations with the bound on $nR_2$ in \eqref{proof-epied2R1R2-bounds}, we obtain
\begin{equation}
\begin{split}
R_2 \leq & \ \logfn{\cov{Y_{1G}}2^{-2R_1} - \sigma_1^2 } - \logfn{\cov{S_{2G}|X_{2G}}} \\[2mm]
& ~~~~~~~~~~~~~+ \logfn{\cov{Y_{2G}|S_{2G}}} - \logfn{h_{21}^22^{2R_1} + \sigma_2^2} \\[2mm]
= & \; \logfn{\frac{\cov{Y_{2G}|S_{2G}}}{\cov{S_{2G}|X_{2G}}}} + \logfn{\frac{\cov{Y_{1G}}2^{-2R_1} - \sigma_1^2}{h_{21}^22^{2R_1} + \sigma_2^2} } .
\end{split}
\end{equation}
\end{proof}
\subsection{Numerical Results}
\begin{figure}
\centering
\includegraphics[width=0.7\textwidth]{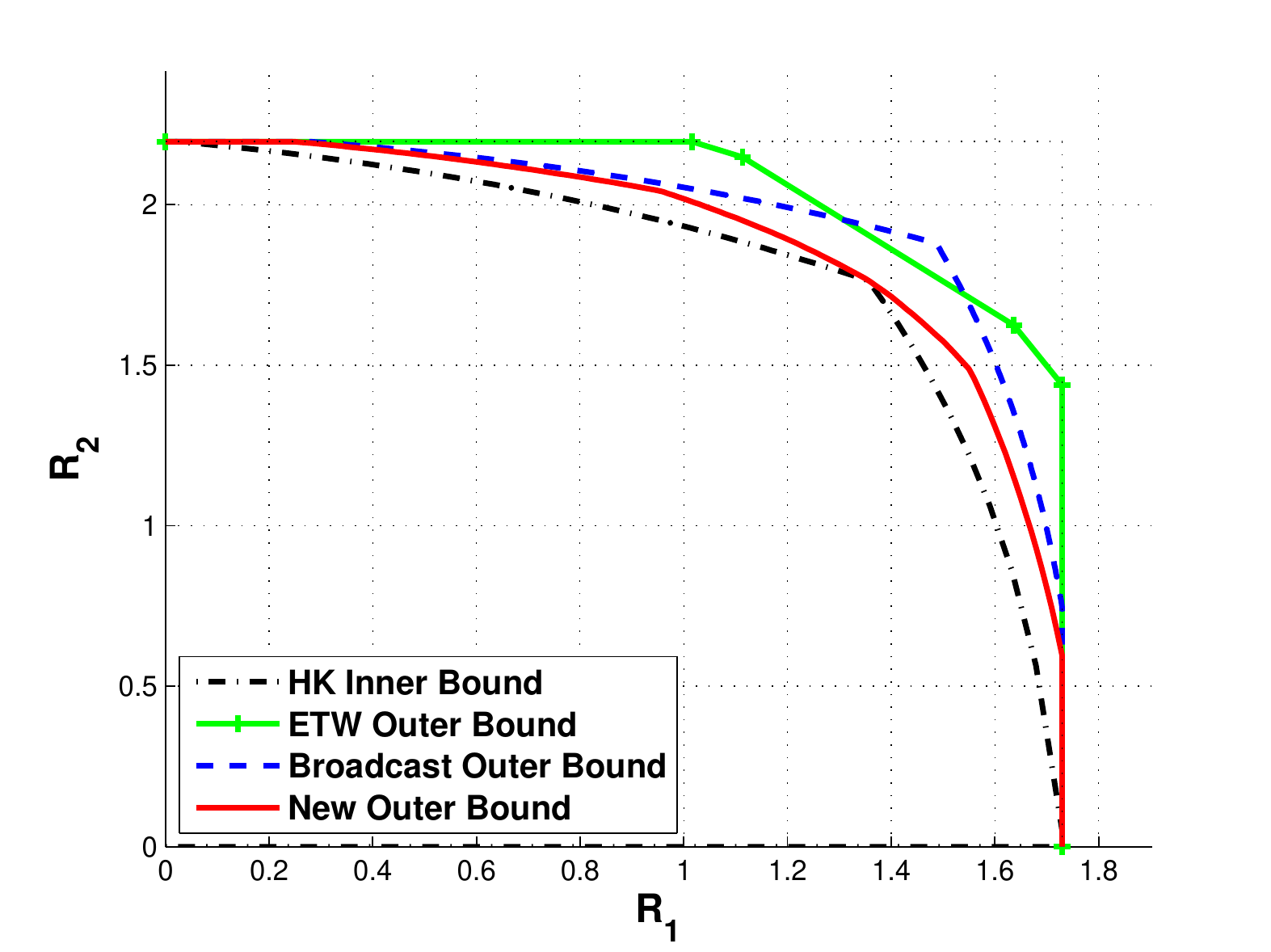}
\caption{Two user Gaussian interference channel ($P_1 = 10, P_2 = 20, h_{12}^2 = 0.04, h_{21}^2 = 0.09$) in low interference regime: Bounds on the capacity region.}
\label{fig:Bounds_Region_cmp-shang}
\end{figure}
\begin{figure}
\centering
\includegraphics[width=0.7\textwidth]{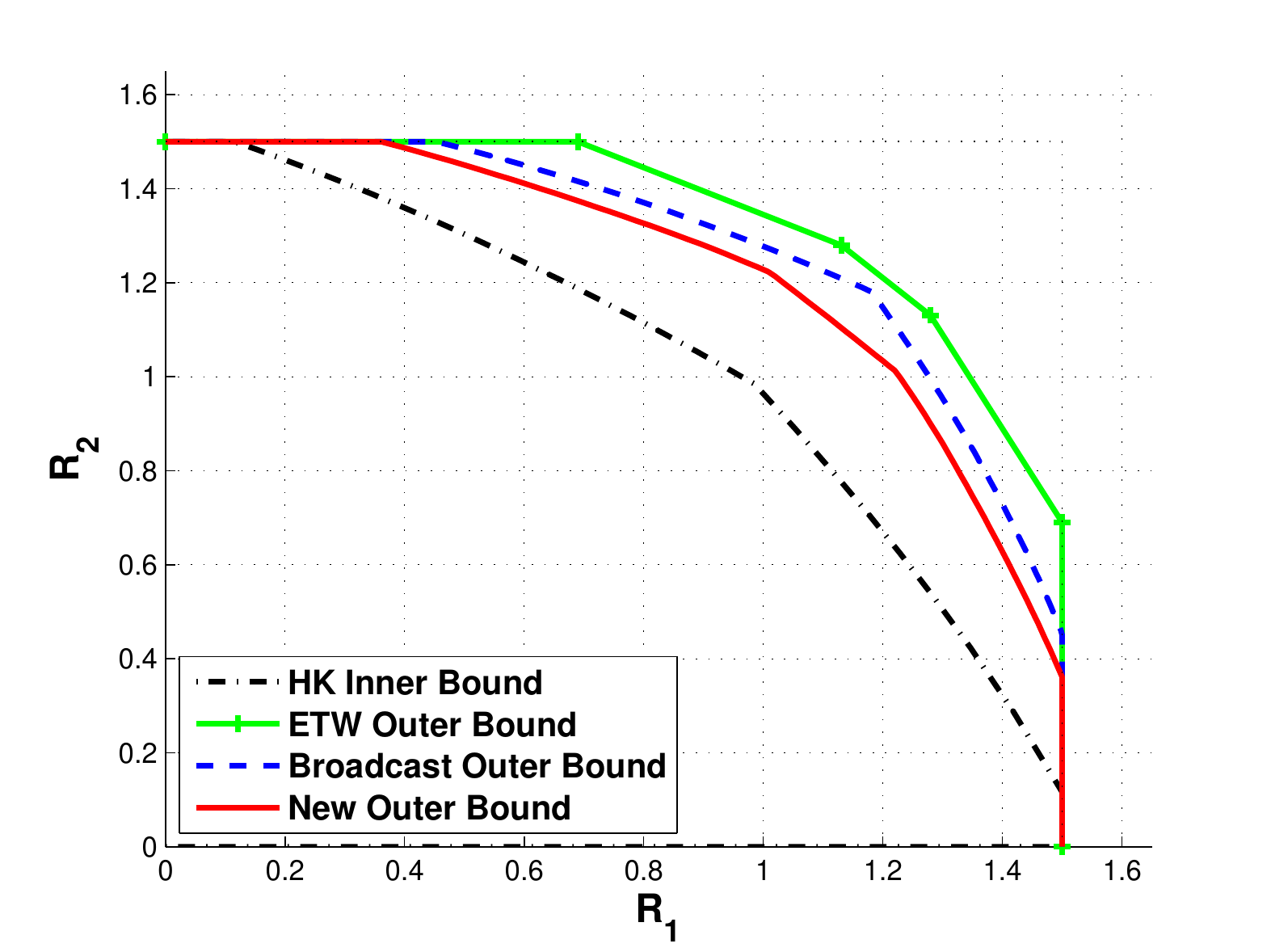}
\caption{Two user symmetric Gaussian interference channel ($P = 7, h^2 = 0.2$): Bounds on the capacity region.}
\label{fig:Bounds_Region_cmp-motahari}
\end{figure}
In Figures~\ref{fig:Bounds_Region_cmp-shang} and \ref{fig:Bounds_Region_cmp-motahari}, we plot the new outer bound, i.e., EPI-based ETW outer bound, along with the original ETW outer bound and the \BCbound. To compare the outer bounds, we also plot a special case of the Han-Kobayashi inner bound, that does not include time sharing and is limited to only Gaussian distributions for the private and common messages. Since the EPI-based ETW outer bound contains the original ETW outer bound and broadcast channel outer bound as special cases, it is obviously tighter. Figure~\ref{fig:Bounds_Region_cmp-shang} corresponds to $P_1 = 10, P_2 = 20, h_{12}^2 = 0.04, h_{21}^2 = 0.09$, which satisfy the condition \eqref{condition_asym_simple} for low interference, and hence the inner and outer bounds meet at one point to give the sum capacity. Figure~\ref{fig:Bounds_Region_cmp-motahari} corresponds to $P_1 = P_2 = 7, h_{12}^2 = h_{21}^2 = 0.2$, which do not satisfy the condition \eqref{condition_asym_simple} for low interference, and hence inner and outer bounds do not meet.

As discussed in Section~\ref{sec:obersvations}, the outer bounds presented in this paper are tightened versions of the ETW outer bounds, obtained by considering a general class of genie signals and using EPI instead of the worst case noise result. Similar approach has been taken independently by two other groups - Shang, Kramer and Chen \cite{Shang-Kramer-Chen-2007} and Motahari and Khandani \cite{Motahari-Khandani-2007}. The main difference in the approaches is that \cite{Shang-Kramer-Chen-2007} and \cite{Motahari-Khandani-2007} use extremal inequality \cite{Liu-Viswanath-IT2007} instead of EPI. Although the extremal inequalities proposed in \cite{Liu-Viswanath-IT2007} are more general than EPI, both are equivalent for the purpose of this paper. Hence we believe that both the approaches should yield the same bounds. Shang et. al. tightened only the sum rate outer bound \eqref{onebit-bound-etw} and hence their outer bound, equivalent to Lemma~\ref{lemma:Epied-OneBit-Bound}, is weaker compared to Theorem~\ref{Th:Epied-Bound} that includes Lemma~\ref{lemma:Epied-2R1R2-Bound} as well. Motahari et. al. tightened all the ETW outer bounds and hence their outer bound is equivalent to Theorem~\ref{Th:Epied-Bound}. We may compare Figure~\ref{fig:Bounds_Region_cmp-motahari} with Figure~3 in \cite{Motahari-Khandani-2007} and Figure~\ref{fig:Bounds_Region_cmp-shang} with Figure~4 in \cite{Shang-Kramer-Chen-2007}. 

%
\section{Gaussian Interference Network: Sum Capacity in Low Interference Regime}
\label{Sec:network-sumcapacity}
In section~\ref{Sec:twouser-sumcapacity}, we established the sum capacity of the two-user Gaussian interference channel in a low interference regime. The intuition is that if the interference is low enough, the receiver will not able to exploit the structure in the interference, and hence treating interference as noise achieves the sum capacity. It is natural to verify if the result can be extended to an arbitrary interference network, and if it does, to see how the interference threshold scales with the number of users. In this section, we first consider two special cases of the general interference network: the {\em many-to-one interference channel}, where only one user experiences interference, and the {\em one-to-many interference channel}, where the interference is generated by only one user. For these two special cases, we use a genie similar to that used for the two-user interference channel, which we call now a scalar genie, to propose conditions under which treating interference as noise achieves the sum capacity. 

Using the scalar genie, Shang et. al. derived conditions for the optimality of treating interference as noise for an arbitrary Gaussian interference network \cite{Shang-Kramer-Chen-ISIT2008}. For symmetric interference channels, this results in an INR$_{\textrm{total}}$ threshold, below which treating interference as noise achieves sum capacity, that is independent of the number of users. Here we use the notation INR$_{\textrm{total}}$ for  a symmetric interference channel to denote the total interference-to-noise ratio. We show that there exists an alternative construction of the genie, where each receiver is provided with multiple genie signals, resulting in a INR$_{\textrm{total}}$ threshold for the symmetric three-user interference channel, that is higher than the INR threshold for the symmetric two-user interference channel.
\subsection{Many-to-one and One-to-many interference channels}
The \manyone and \onemany interference channels are studied in \cite{Alex2007,Bresler2007}, where the capacity region is characterized to within a constant number of bits.\\
\textit{Many-to-one:}
In a \manyone \gic only one user experiences the interference, i.e.,
\[
h_{rt} = 0, \forall t \neq r, \forall r \neq 1
\]
where we assume that the user $1$ is the unlucky user without any loss of generality. Thus the \manyone \gic is parameterized by $\{P_1,P_2,\cdots,P_M,h_{12},h_{13},\cdots,h_{1M}\}$:
\begin{equation}
\begin{split}
Y_1 = & X_1 + \sum_{t = 2}^{M} h_{1t}X_t + Z_1 \\
Y_r = & X_r + Z_r, \text{ for } r = 2, 3, \cdots, M .
\label{many_to_one_ic}
\end{split}
\end{equation}
\smallskip
\textit{One-to-many:}
In a \onemany \gic only one user causes the interference, i.e.,
\[
h_{rt} = 0, \forall r \neq t, \forall t \neq 1
\]
where we assumed that user $1$ is the interfering user. Thus the \onemany \gic is parameterized by $\{P_1,P_2,\cdots,P_M,h_{21},h_{31},\cdots,h_{M1}\}$:
\begin{equation}
\begin{split}
Y_1 = & X_1 + Z_1 \\
Y_r = & h_{r1}X_1 + X_r + Z_r, \text{ for } r = 2, 3, \cdots, M .
\label{one_to_many_ic}
\end{split}
\end{equation}
\begin{theorem} \label{TH:sumcapacity-manyone}
For a many-to-one interference channel \eqref{many_to_one_ic} satisfying
\begin{equation}
\sum_{i = 2}^{M} h_{1i}^2 \leq 1
\label{condition_sumcapacity_manyone}
\end{equation}
treating interference as noise achieves the sum capacity, which is given by
\begin{equation}
\Csum = \frac{1}{2}\log\left(1 + \frac{P_1}{\sum_{i = 2}^{M}h_{1i}^2P_i}\right) + \frac{1}{2}\sum_{i = 2}^{M}\log\left(1 + P_i\right).
\label{sumcapacity_manyone}
\end{equation}
\end{theorem}
\medskip
\begin{proof}
Allowing the interfering users to cooperate can only increase the sum capacity. Let $\uy_I$ be the vector denoting the collective received signal, $\ux_I$ and $\uz_I$ denote the corresponding transmit and noise vectors and $\uh = [h_{12} \ h_{13} \ \cdots \ h_{1M}]^T$ to arrive at
\[
\begin{split}
Y_1 = & \ X_1 + \uh^T\ux_I + Z_1 \\
Y_I = & \ \ux_I + \uz_I
\end{split}
\]
Let $S_I = \uh^T\ux_I + W_I$ be the side information given to the (collective) receivers of the interfering users. Here $W_I$ is zero mean, unit variance, Gaussian random variable. Using Fano's inequality, we have
\[
\begin{split}
n\sum_{i=1}^{M}(R_i - \epsilon_n) \leq & \ I(X_1^n;Y_1^n) + I(\ux_I^n;\uy_I^n,S_I^n) \\
= & \ I(X_1^n;Y_1^n) + I(\ux_I^n;S_I^n)  + I(\ux_I^n;\uy_I^n|S_I^n)\\
= & \ \entropy(Y_1^n) - \entropy(S_I^n) + \entropy(S_I^n) - \entropy(W_I^n)  + \entropy(\uy_I^n|S_I^n) - \entropy(\uz_I^n|W_I^n)\\
= & \ \entropy(Y_1^n) - n\entropy(W_{IG})  + \entropy(\uy_I^n|S_I^n) - n\entropy(\uz_{IG}|W_{IG})\\
\stackrel{(a)} \leq & \ n\entropy(Y_{1G}) - n\entropy(W_{IG}) + n\entropy(\uy_{IG}|S_{IG}) - n\entropy(\uz_{IG}|W_{IG})\\
= & \ nI(X_{1G};Y_{1G}) + nI(\ux_{IG};\uy_{IG},S_{IG})
\end{split}
\]
where the step (a) follows from Lemma~\ref{lemma-concavity-conditonal-entropy}. Thus the genie is useful. If \eqref{condition_sumcapacity_manyone} is true, then the random variable $W_I$ can be chosen such that
\[
W_{I} = \uh^T\uz_I + V
\]
where the Gaussian random variable $V$ is independent of $\uz_{I}$. Therefore,
\[
S_{I} = \uh^T\uy_I + V 
\]
and hence $I(\ux_{IG};\uy_{IG},S_{IG}) = I(\ux_{IG};\uy_{IG})$ making the genie smart. Hence the theorem follows.
\end{proof}
\smallskip
\begin{theorem} \label{TH:sumcapacity-onemany}
For a one-to-many interference channel \eqref{one_to_many_ic} satisfying
\begin{equation}
\sum_{i = 2}^{M} \frac{h_{i1}^2P_1 + h_{i1}^2}{h_{i1}^2P_1 + 1} \leq 1
\label{condition_sumcapacity_onemany}
\end{equation}
treating interference as noise achieves the sum capacity, which is given by
\begin{equation}
\Csum = \frac{1}{2}\log\left(1 + P_1\right) + \frac{1}{2}\sum_{i = 2}^{M}\log\left(1 + \frac{P_i}{h_{i1}^2P_1 + 1}\right)
\label{sumcapacity_onemany}
\end{equation}
\end{theorem}
\smallskip
\begin{proof}
We prove this theorem directly without the aid of a genie.
\[
\begin{split}
n(\Csum - M\epsilon_n) \leq & \ I(X_1^n;Y_1^n) + \sum_{i = 2}^{M}I(X_i^n;Y_i^n) \\
= & \ \entropy(Y_1^n) - \entropy(Y_1^n|X_1^n) +  \sum_{i = 2}^{M} \entropy(Y_i^n) - \entropy(Y_i^n|X_i^n) \\
= & \ \entropy(Y_1^n) - n\entropy(Y_{1G}|X_{1G}) +  \sum_{i = 2}^{M} \entropy(Y_i^n) - \entropy(Y_i^n|X_i^n) \\
\leq & \ \entropy(Y_1^n) - n\entropy(Y_{1G}|X_{1G}) +  \sum_{i = 2}^{M} n\entropy(Y_{iG}) - \entropy(Y_i^n|X_i^n)
\end{split}
\]
To finish the proof, we further need to show that
\[
\begin{split}
& \ \entropy(Y_1^n) - \sum_{i = 2}^{M} \entropy(Y_i^n|X_i^n) \\
= & \ \entropy(X_1^n + Z_1^n) - \sum_{i = 2}^{M} \entropy(h_{i1}X_1^n + Z_i^n) \\
= & \ \sum_{i = 2}^{M} \lambda_i \entropy(X_1^n + Z_1^n) - \entropy(h_{i1}X_1^n + Z_i^n)
\end{split}
\] is maximized by $X_1^n = X_{1G}^n$, for some $\{ \lambda_i \}_{i = 2}^{M}$ such that $\sum_{i = 2}^{M} \lambda_i = 1$. If  \eqref{condition_sumcapacity_onemany} holds, it is possible to chose $\{ \lambda_i \}_{i = 2}^{M}$ satisfying
\[
\lambda_i \geq \frac{h_{i1}^2P_1 + h_{i1}^2}{h_{i1}^2P_1 + 1}.
\]
For this choice of $\lambda_{i}$, from Lemma~\ref{new-extremal-inequality}, it follows that $\lambda_i\entropy(h_{i1}X_1^n + h_{i1}Z_1^n) - \entropy(h_{i1}X_1^n + Z_i^n)$, and therefore $\lambda_i\entropy(X_1^n + Z_1^n) - \entropy(h_{i1}X_1^n + Z_i^n)$, is maximized when $X_1^n = X_{1G}^n$. Hence the result follows.
\end{proof}
\smallskip
\begin{remark}
Theorems~\ref{TH:sumcapacity-manyone} and \ref{TH:sumcapacity-onemany} can be shown to special cases of Theorem~4 in \cite{Shang-Kramer-Chen-ISIT2008}.
\end{remark}
\subsection{Vector genie}
We now propose a systematic construction of an useful genie for an arbitrary interference network. We call this a vector genie because it involves giving multiple side information signals to each receiver. This vector genie can be thought of as a generalization of the ETW genie \eqref{genie_etw} developed for the two-user interference channel. We need to define an ordering function before constructing the vector genie signal. 
\begin{definition}[Ordering function]
We call a function $\pi: \{1,2,\cdots,M\} \rightarrow \{1,2,\cdots,M\}$ an ordering function if it satisfies the following properties
\begin{equation}
\begin{split}
\{1,\pi(1),\pi^{(2)}(1),\cdots,\pi^{(M-1)}(1)\} = & \ \{1,2,\cdots,M\} \\
\pi^{(M)}(r) = & \ r, \forall r
\end{split}
\label{property-ordering-function}
\end{equation}
where $\pi^{(j)}(.)$ denotes the function $\pi(.)$ operated $j$ times. 
\end{definition}
\begin{definition}
Suppose $Y_{r}$ is a random variable that is an affine  combination of the variables $\{X_{t}\}_{t=1}^{M}$. For any $\mathcal{A} \subseteq \{1,2,\cdots,M\}$, $Y_{r}\backslash{}\{X_{t}, t \in \mathcal{A}\}$ denotes the random variable obtained after removing the contributions of $\{X_{t}, t \in \mathcal{A}\}$ from $Y_{r}$.
\end{definition}
For any fixed ordering function $\pi$, let \[\us_r = [S_{r,1} \ S_{r,2} \ \cdots S_{r,M-1}]^{\top}\] be the side information given to the receiver $r$, defined as
\begin{equation}
S_{r,k} = Y_{\pi^{(k)}(r)}\backslash \{X_{\pi^{(j)}(r)}\}_{j=1}^{k}, \text{ for } k = 1,2,\cdots,M-1.
\label{asymptotic-vector-genie}
\end{equation}
For example, consider the three user interference network. With the ordering function \[\pi(1) = 2, \pi(2) = 3, \pi(3) = 1\] we see that the genie signals defined by (\ref{asymptotic-vector-genie}) are:
\begin{displaymath}
\begin{array}{|c|c|c|c|}
\hline
       		& r = 1 & r = 2 & r = 3 \\
\hline
Y_r :  & X_1 + h_{12}X_2 + h_{13}X_3 + Z_{1} 	& X_2 + h_{21}X_1 + h_{23}X_3 + Z_{2} 	& X_3 + h_{31}X_1 + h_{32}X_2 + Z_{3} \\
S_{r,1}:  	& h_{21}X_1 + h_{23}X_3 + Z_{2}		& h_{32}X_2 + h_{31}X_1 + Z_{3}		& h_{13}X_3 + h_{12}X_2 +  Z_{1}\\
S_{r,2}: 	& h_{31}X_1 + Z_{3}			& h_{12}X_2 + Z_{1}			& h_{23}X_3 + Z_{2}\\
\hline
\end{array}
\end{displaymath}
\medskip
The following properties of the genie \eqref{asymptotic-vector-genie} are useful in deriving the outer bounds.
\begin{prop}
For each $r$, the genie signal $S_{r,M-1}$ is interference free, i.e., $S_{r,M-1}\backslash{}X_r$ is Gaussian.
\label{vector-genie-asym-prop1}
\end{prop}
\begin{proof}
From the construction of the genie (\ref{asymptotic-vector-genie}), we have
\[
S_{r,M-1} =  \ Y_{\pi^{(M-1)}(r)}\backslash{}\{X_{\pi^{(j)}(r)}\}_{j=1}^{M-1}
\]
which implies that
\[
\begin{split}
S_{r,M-1}\backslash{}X_r \stackrel{(a)} = & \ Y_{\pi^{(M-1)}(r)}\backslash{}\{X_{\pi^{(j)}(r)}\}_{j=1}^{M} \\
   \stackrel{(b)} = & \ Y_{\pi^{(M-1)}(r)}\backslash{}\{X_j\}_{j=1}^{M} \\
                  = & \ Z_{\pi^{(M-1)}(r)}
\end{split}
\]
where steps (a) and (b) follow from the property \eqref{property-ordering-function} of the ordering function $\pi$.
\end{proof}
\begin{prop}
For each receiver $r$, define
\begin{equation}
\uyt_r = [Y_r \ S_{r,1} \ S_{r,2} \ \cdots S_{r,M-2}]
\label{definition-ytilde}
\end{equation}
then
\[
\us_r = \uyt_{\pi(r)}\backslash{}X_{\pi(r)} .
\]
\label{vector-genie-asym-prop2}
\end{prop}
\begin{proof}
The result follows because \[S_{r,1} = Y_{\pi(r)}\backslash{}X_{\pi(r)}\] and for $k = 2,3,\cdots,M-1$,
\[
\begin{split}
S_{r,k} = & \ Y_{\pi^{(k)}(r)}\backslash{}\{X_{\pi^{(j)}(r)}\}_{j=1}^{k} \\
= & \ Y_{\pi^{(k-1)}(\pi(r))}\backslash{}\{X_{\pi^{(j)}(r)}\}_{j=1}^{k} \\
= & \ Y_{\pi^{(k-1)}(\pi(r))}\backslash{}\left\{ \{X_{\pi^{(j)}(r)}\}_{j=2}^{k},X_{\pi(r)}\right\} \\
= & \ Y_{\pi^{(k-1)}(\pi(r))}\backslash{}\left\{ \{X_{\pi^{(j)}(\pi(r))}\}_{j=1}^{k-1},X_{\pi(r)}\right\} \\
= & \ S_{\pi(r),k-1}\backslash{}X_{\pi(r)}.
\end{split}
\]
\end{proof}
We now proceed to show that the vector genie \eqref{asymptotic-vector-genie} is useful and derive an outer bound on the sum capacity.
\begin{theorem}
For any ordering function $\pi$, the genie defined in \eqref{asymptotic-vector-genie} is useful, i.e., the sum capacity of the interference network (\ref{interference_network}) is upper bounded by
\[
\Csum \leq \sum_{i=1}^{M} I(X_{iG};Y_{iG},\us_{iG})
\]
where the genie signals $\{\us_{i}\}$ are defined in (\ref{asymptotic-vector-genie}). 
\label{asym-sumcapacity-interference-network}
\end{theorem}
\begin{proof}
\[
\begin{split}
n(\Csum - M\epsilon_n)
\leq & \ \sum_{i=1}^{M} I\left(X_i^n;Y_i^n,\us_i^n\right) \\
=    & \ \sum_{i=1}^{M} \entropy\left(Y_i^n,\us_i^n\right) - \entropy\left(Y_i^n,\us_i^n|X_i^n\right) \\
\stackrel{(a)} =    & \ \sum_{i=1}^{M} \entropy\left(Y_i^n,\us_i^n\right) - \entropy\left(\uyt_i^n,S_{i,M-1}^n|X_i^n\right) \\
=    & \ \sum_{i=1}^{M} \entropy\left(\us_i^n\right) + \entropy\left(Y_i^n|\us_i^n\right) - \entropy\left(S_{i,M-1}^n|X_i^n\right) - \entropy\left(\uyt_i^n|S_{i,M-1}^n,X_i^n\right) \\
=    & \ \sum_{i=1}^{M} \entropy\left(Y_i^n|\us_i^n\right) - \entropy\left(S_{i,M-1}^n|X_i^n\right) + \sum_{i=1}^{M} \entropy\left(\us_i^n\right) - \entropy\left(\uyt_i^n|S_{i,M-1}^n,X_i^n\right) \\
\stackrel{(b)} =    & \ \sum_{i=1}^{M} \entropy\left(Y_i^n|\us_i^n\right) - \entropy\left(S_{i,M-1}^n|X_i^n\right) + \sum_{i=1}^{M} \entropy\left(\us_i^n\right) - \entropy\left(\us_{\pi^{(M-1)}\left(i\right)}^n\right)~~~~~\\
 \stackrel{(c)} =    & \ \sum_{i=1}^{M} \entropy\left(Y_i^n|\us_i^n\right) - \entropy\left(S_{i,M-1}^n|X_i^n\right) \\
\stackrel{(d)} =    & \ \sum_{i=1}^{M} \entropy\left(Y_i^n|\us_i^n\right) - n\entropy\left(S_{iG,M-1}|X_{iG}\right) \\
\end{split}
\]
\[
\begin{split}
\hphantom{\Csum - M\epsilon_n} \stackrel{(e)} \leq & \ \sum_{i=1}^{M} n\entropy\left(Y_{iG}|\us_{iG}\right) - n\entropy\left(S_{iG,M-1}|X_{iG}\right) \\
\end{split}
\]
where step (a) follows from the definition of $\uyt_r$ \eqref{definition-ytilde}, step (b) follows from Propositions~\ref{vector-genie-asym-prop1} and \ref{vector-genie-asym-prop2}, step (c) follows because $\displaystyle \{\pi^{(M-1)}(i)\}_{i=1}^M = \{1,2,\cdots,M \}$, step (d) follows from Proposition~\ref{vector-genie-asym-prop1}, and finally step (e) follows from Lemma~\ref{lemma-concavity-conditonal-entropy}. We have shown that $\{X_{iG}^n\}_{i=1}^M$ maximizes $\sum_{i=1}^{M} I(X_i^n;Y_i^n,\us_i^n)$ and clearly the maximum is given by $n\sum_{i=1}^{M} I(X_{iG};Y_{iG},\us_{iG})$, and hence we have the result.
\end{proof}
\begin{remark}
The vector genie is a generalization of the ETW genie and hence Theorem~\ref{asym-sumcapacity-interference-network} simplifies to the ETW bound \eqref{onebit-bound-etw} for the two-user interference channel. For the two-user interference channel, the ETW genie is also used to derive outer bounds (\ref{R1-bound-etw}-\ref{R1-2R2-bound-etw}) on the entire capacity region. In a similar fashion, the vector genie can also be used to derive outer bounds on the entire capacity region of an arbitrary Gaussian interference network. 
\end{remark}

Similar to the two-user case, we proceed to tighten the outer bound by correlating the noise terms in the genie signals to the receiver noise. In particular, we explore if there exists a genie that is not just useful, but also smart, to establish the sum capacity in the low interference regime.

\subsection{Three user symmetric interference channel}
To simplify the presentation, we will restrict our attention to the symmetric three user channel, i.e., $P_{t} = P, \forall t$ and $h_{rt} = h, \forall r \neq t$. To make the genie smart, we let the noise terms in the genie signals be correlated to the noise at the receiver.

\begin{displaymath}
\begin{array}{|c|c|c|c|}
\hline
       		& r = 1 & r = 2 & r = 3 \\
\hline
Y_r :  & X_1 + hX_2 + hX_3 + Z_{1} 	& X_2 + hX_1 + hX_3 + Z_{2} 	& X_3 + hX_1 + hX_2 + Z_{3} \\
S_{r,1}:  	& hX_1 + hX_3 + h\eta_1W_{11}		& hX_2 + hX_1 + h\eta_1W_{21}		& hX_3 + hX_2 + h\eta_1W_{31}\\
S_{r,2}: 	& hX_1 + h\eta_2W_{12}			& hX_2 + h\eta_2W_{22}			& hX_3 + h\eta_2W_{32}\\
\hline
\end{array}
\end{displaymath}
\medskip

\noindent Here $\{ W_{rk} \}_{r=1,k=1}^{3,2}$ are zero mean, unit variance, Gaussian random variables, and $\eta_1,\eta_2$ are real variables. Let $\Sigma$ denote the covariance matrix of the random vector $[Z_r \ W_{r1} \ W_{r2}]^\top$ (which is independent of $r$):
\begin{equation}
\Sigma = \left[\begin{array}{c c c}
1 & \rho_1 & \rho_2 \\
\rho_1 & 1 & \rho_{12} \\
\rho_2 & \rho_{12} & 1
\end{array}\right]
\end{equation}
Thus the genie is parameterized by $\{\Sigma,\eta_1,\eta_2\}$.
\begin{lemma}[Useful Genie]
The genie is useful i.e.,
\[
\Csum \leq \sum_{i=1}^{M}I(X_{iG};Y_{iG},\us_{iG})
\]
when
\begin{equation}
 \cov{[Z_1 \ h\eta_1W_{11}]^\top|W_{12}} -  \cov{[h\eta_1W_{11} \ h\eta_2W_{12}]^\top} \succcurlyeq 0 .
 \label{condition_usefulgenie_3user}
\end{equation}
\end{lemma}
\begin{proof}
Following the proof of Theorem~\ref{asym-sumcapacity-interference-network}, we only need to show that
\[
\begin{split}
\sum_{i=1}^{M} \entropy(\us_i^n) - \entropy(\uyt_i^n|S_{i,M-1}^n,X_i^n) =
& \ \sum_{i=1}^{M} \entropy(\us_i^n) - \entropy\left(\uyt_{\pi(i)}^n|S_{\pi(i),M-1}^n,X_{\pi(i)}^n\right) \\
\end{split}
\]
is maximized by $\{X_{iG}^n\}$. For $i = 1$,
\begin{equation} \label{eq:proof_usefulgenie_3user}
\begin{split}
\entropy(\us_i^n) - & \ \entropy\left(\uyt_{\pi(i)}^n|S_{\pi(i),M-1}^n,X_{\pi(i)}^n\right) = \entropy(\us_1^n) - \entropy\left(\uyt_2^n|S_{2,2}^n,X_{2}^n\right) \\
& \ \entropy\left(\left[ \begin{array}{c} hX_1^n + hX_3^n + h\eta_1W_{11}^n \\ hX_1^n + h\eta_2W_{12}^n \end{array} \right]\right) - \ \entropy\left(\left[ \begin{array}{c} hX_1^n + hX_3^n + Z_2^n \\ hX_1^n + h\eta_1W_{21}^n \end{array} \right]|W_{22}\right).
\end{split}
\end{equation}
Using Lemmas~\ref{lemma-conditonal-entropy-mmse} and \ref{lemma-worstcase-noise-vector}, it follows that \eqref{eq:proof_usefulgenie_3user} is maximized by $\{X_{iG}^n\}$ if the condition \eqref{condition_usefulgenie_3user} holds.
\end{proof}
We next give the conditions for the genie to be smart in the following lemma, which is an extension of Lemma~\ref{smart_genie}.
\begin{lemma}[Smart Genie]
The genie is smart, i.e., 
\begin{equation} \label{expression_smartgenie_3user}
I(X_{iG};Y_{iG},\us_{iG}) = I(X_{iG};Y_{iG})
\end{equation}
 iff the following conditions hold
\begin{equation}
\begin{split}
\eta_1\rho_1 = & \ 1 + 2h^2P - hP \\
\eta_2\rho_2 = & \ 1 + 2h^2P.
\end{split}
\label{condition_smartgenie_3user}
\end{equation}
\end{lemma}
\smallskip
\begin{proof}
Since \[I(X_{iG};Y_{iG},\us_{iG}) = I(X_{iG};Y_{iG}) + I(X_{iG};\us_{iG}|Y_{iG})\] \eqref{expression_smartgenie_3user} is equivalent to 
\begin{equation} \label{eq1-smartgenie-3user-proof}
I(X_{iG};\us_{iG}|Y_{iG}) = 0.
\end{equation} 
From Lemma~\ref{lemma-markov-chain2}, it follows that \eqref{eq1-smartgenie-3user-proof} is true iff
\begin{equation} \label{eq2-smartgenie-3user-proof}
\begin{split}
I(X_{iG};S_{i,1G}|Y_{iG}) = & \ 0 \\
I(X_{iG};S_{i,2G}|Y_{iG}) = & \ 0.
\end{split}
\end{equation}
Using Lemma~\ref{lemma-markov-chain1}, we have
\[
\begin{split}
I(X_{1G};S_{1,1G}|Y_{1G}) = & \ 0 \\
\iff I(X_{1G};X_{1G}+X_{3G} + \eta_{1}W_{11}|X_{1G} + hX_{2G} + hX_{3G} + Z_{1}) = & \ 0 \\
\iff \expect{(X_{3G} + \eta_{1}W_{11})(hX_{2G} + hX_{3G} + Z_{1})}  = & \ \expect{( hX_{2G} + hX_{3G} + Z_{1})^{2}} \\
\iff hP + \eta_{1}\rho_{1}  = & \ 1 + 2h^{2}P.
\end{split}
\]
and
\[
\begin{split}
I(X_{1G};S_{1,2G}|Y_{1G}) = & \ 0 \\
\iff I(X_{1G};X_{1G}+ \eta_{2}W_{12}|X_{1G} + hX_{2G} + hX_{3G} + Z_{1}) = & \ 0 \\
\iff \expect{\eta_{2}W_{12}(hX_{2G} + hX_{3G} + Z_{1})}  = & \ \expect{(hX_{2G} + hX_{3G} + Z_{1})^{2}} \\
\iff \eta_{2}\rho_{2}  = & \ 1 + 2h^{2}P.
\end{split}
\]
\end{proof}
\smallskip
\begin{theorem}
\label{prop:sumcapacity-threeuser}
For the symmetric three user Gaussian interference channel, suppose there exist $\{\Sigma \succcurlyeq 0, \eta_1, \eta_2 \}$ satisfying \eqref{condition_usefulgenie_3user} and \eqref{condition_smartgenie_3user}, then treating interference as noise achieves the sum capacity, which is given by
\[
\Csum = \frac{3}{2}\log\left(1 + \frac{P}{1+2h^2P}\right).
\]
\end{theorem}
\bigskip
Unlike in the two-user case, we have not been able to provide an explicit equation for the threshold on $h$ (as a function of $P$) below which treating interference as noise achieves the sum capacity. Nevertheless, for every $P$, admissible values of $h$ can be found numerically by searching for the parameters $\{\Sigma \succcurlyeq 0, \eta_1, \eta_2 \}$ that satisfy the conditions in Theorem~\ref{prop:sumcapacity-threeuser}. 

\begin{figure}
\centering
\includegraphics[width=0.7\textwidth]{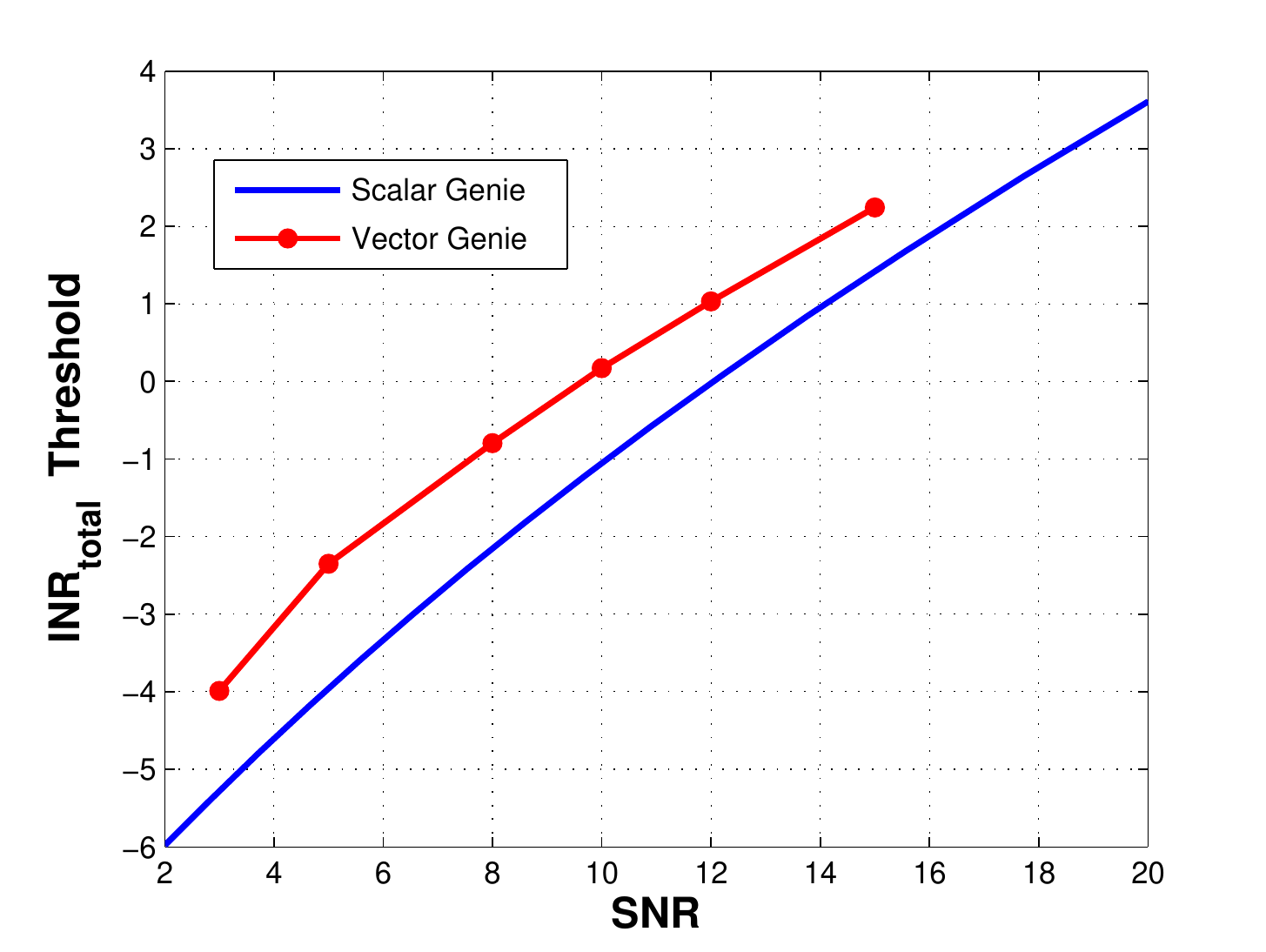}
\caption{Three user symmetric Gaussian interference channel: INR$_{\textrm{total}}$ threshold, below which treating interference as noise achieves the sum capacity, as a function of SNR.}
\label{fig:SNRvsINR-threeuser}
\end{figure}

Using a scalar genie similar to that used for the two-user interference channel, Shang et. al. obtained a threshold on INR$_{\textrm{total}}$ that is independent of the number of users \cite[Theorem~4]{Shang-Kramer-Chen-ISIT2008}. In Figure~\ref{fig:SNRvsINR-threeuser}, we plot a few admissible points that are computed numerically along with the INR$_{\textrm{total}}$ obtained using the scalar genie. An increase of more than $1$ dB in the $\text{INR}_{\text{total}}$ threshold is seen by using the vector genie instead of the scalar genie. Note that  $\text{INR}_{\text{total}}$ threshold obtained using the vector genie for the three-user interference channel is greater than the INR threshold for the two-user interference channel (which is same as the $\text{INR}_{\text{total}}$ threshold obtained using the scalar genie).

Although, the thresholds we obtain in this paper are only lower bounds to the optimal threshold, we believe that the trend shown by the vector genie holds true, i.e., {\em the optimal interference threshold, below which treating interference as noise achieves sum capacity, increases with the number of users}. The optimality of treating interference as noise in the low  interference regime implies that the receivers are not able to exploit the structure in the interference. With more users in the network, the ability of the receiver to exploit the structure in each of the interfering user's signal can only decrease because the interfering users' signals interfere with each other. 
\section{Conclusions} \label{sec:concl}

We provided new, improved genie-aided outer bounds on the capacity region of a two-user Gaussian interference channel. Using these outer bounds, we showed that treating interference as noise achieves the sum capacity in a low interference regime. Similar results were established in parallel by Shang, Kramer and  Chen \cite{Shang-Kramer-Chen-2007}, and Motahari and Khandani \cite{Motahari-Khandani-2007}. Although the interference threshold, below which treating interference as noise achieves sum capacity, is identical in the three works, the mathematical approach is considerably different. It is also  to be noted that what has  been obtained in all three works is only  a lower bound on the interference threshold, and the question still remains as to  what the {\em optimal interference threshold} is.

A natural extension of the two-user results is the generalization of the optimality of treating interference as noise in the low interference regime to Gaussian interference networks with more than two users. We provided closed form expressions that characterize the low interference regime for the many-to-one and one-to-many interference channels. Furthermore, we generalized the ETW genie \cite{OneBit2007}  to an arbitrary Gaussian interference network, i.e., proposed a systematic  construction of a genie such that treating interference as noise with Gaussian inputs achieve the sum capacity of the genie-aided network. We called this genie a vector genie, because it involves  giving multiple side information signals to each receiver. Similar to \cite{OneBit2007,TelatarTse2007}, this vector genie can be used to derive outer bounds on the entire capacity region.



By correlating the noise terms in the vector genie, we showed that the outer bound can be further tightened to establish the sum capacity in a low interference regime. For reasons of computational complexity, we only considered a three user symmetric interference channel, for which we demonstrated that the total interference threshold can be higher than that for the two-user case. The interesting question that remains  to be answered is: {\em how does the optimal interference threshold scale as a function of the number of interferers in the network?}
\bibliographystyle{IEEE}
\bibliography{2008IT-Annapureddy-Veeravalli-GIC}

\end{document}